\documentclass[10pt, nofootinbib]{iopart}

\ioptwocol
\eqnobysec

\usepackage[numbers,sort&compress]{natbib}

\makeatletter
\let\@fnsymbol\@arabic
\makeatother

\expandafter\let\csname equation*\endcsname\relax
\expandafter\let\csname endequation*\endcsname\relax

\usepackage{amsmath}
\usepackage{amssymb}
\usepackage{bm}          
\usepackage{braket}
\usepackage[usenames, dvipsnames]{color}
\usepackage{dcolumn}     
\usepackage{diagbox} 
\usepackage{epsfig}
\usepackage{graphicx}
\usepackage{hhline}
\usepackage{mathrsfs}
\usepackage{newtxmath, newtxtext}
\usepackage{slashed}
\usepackage{subfigure}

\usepackage[pagebackref=false, colorlinks=true]{hyperref}
\definecolor{reddish}{rgb}{0.7,0.2,0.0}  
\definecolor{blueish}{rgb}{0.1,0.1,1}
\hypersetup{
linkcolor=reddish,      
citecolor=blueish,      
urlcolor=magenta}       

\begin{document}

\title[Matter in Nonvacuum Black Hole Spacetimes]{Self-Gravitating Matter in Stationary and Axisymmetric Black Hole Spacetimes}

\author{
Prashant Kocherlakota$^{\star, \dagger}$%
and
Ramesh Narayan$^{\dagger, \star}$
}

\address{
$^\star$Black Hole Initiative at Harvard University,\\ 
20 Garden St., Cambridge, MA 02138, USA\\
$^\dagger$Center for Astrophysics, Harvard \& Smithsonian,\\
60 Garden St., Cambridge, MA 02138, USA
}

\ead{pkocherlakota@fas.harvard.edu}

\begin{abstract}
All black holes (BHs) in nature are expected to be described by the Kerr vacuum solution of general relativity (GR). However, the Kerr BH interior contains several problematic features such as a Cauchy horizon, a curvature singularity, and a causality-violating region. Non-Kerr BH models, which are used to examine the genericity of these features, typically contain nontrivial matter content. When such self-gravitating matter is minimally-coupled to Einstein-Hilbert gravity, the Einstein equations can be directly used to investigate its physical properties. We examine here the properties of matter in a broad class of stationary and axisymmetric, geodesically-integrable BH spacetimes, and how they are linked to various features of the spacetime geometry. In these spacetimes, we find the matter to typically flow along timelike Killing orbits in the BH exterior, usually exhibiting differential rotation but sometimes additionally also non-rigid rotation. At a horizon, the matter rest-frame energy density, $\epsilon$, and principal normal pressure, $p_n$, are shown to necessarily satisfy $p_n = -\epsilon$, implying that only specific \textit{types} of matter can thread stationary event horizons (e.g., electromagnetic fields but not massless real scalar fields). Furthermore, we introduce Boyer-Lindquist-like coordinates for the nonstationary regions in the BH interior, which show the matter to be comoving with the interior cosmology. We also obtain simple expressions for the expansions of the ingoing and outgoing zero angular momentum null congruences and comment on the light-focussing behavior of the cosmology. Finally, we verify above results explicitly by working with a representative set of well-known BH spacetimes which contain various types of matter -- scalar fields, electromagnetic fields, anisotropic fluids. Some spacetimes have singularities while others have regular interiors. In the exterior, the matter satisfies the weak energy condition. The framework developed here can be extended to cover more general spacetimes.
\end{abstract}



\section{Introduction} 
\label{sec:SecI_Introduction}

The discovery of the Kerr metric \cite{Kerr1963}, a stationary, vacuum solution to the Einstein equation, was a monumental achievement, realized nearly five decades after the discovery of general relativity (GR; \cite{Einstein1922}). Since the Kerr metric can be used to describe the spacetime geometry of a vacuum, spinning black hole (BH) in GR, it is hypothesized to describe \textit{all} astrophysical BHs in nature. With the first detection of gravitational waves emitted from a binary BH merger \cite{LIGO+2016a} and the first horizon-scale images of supermassive BHs \cite{EHTC+2019a, EHTC+2022a}, we have entered a new era of BH astrophysics, and the Kerr BH metric continues to be of critical importance in interpretations of strong-gravity phenomena. Furthermore, fundamental examinations of the validity of the Kerr metric in describing astrophysical black holes are also swiftly becoming possible \cite{LSC+2019, LSC+2021a, LSC+2021b, Johannsen+2010, EHTC+2019f, Psaltis+2020, EHTC+2022f}.

From a purely theoretical standpoint, however, the interior geometry of a Kerr BH contains several undesirable features such as a Cauchy horizon (which is also unstable due to a blueshift/mass-inflation instability \cite{Poisson+1989, McNamara1978, Sbierski2022}), a spacetime curvature singularity, as well as causality-violating regions that permit closed timelike curves (see, e.g., Ref. \cite{Wiltshire+2009}). While the instability of the Cauchy horizon might render some of these issues physically irrelevant, an alternative approach to finding resolutions involves careful explorations of generic features of non-Kerr BH models.

For instance, several singularity-free ``regular'' BH models have been constructed and the properties of the self-gravitating matter generating these spacetimes have been carefully analyzed \cite{Bardeen1968, Ayon-Beato+1998, Ayon-Beato+2000, Bronnikov2000, Dymnikova2004, Hayward2006, Modesto2008, Caravelli+2010, Bambi+2013, Frolov2016, Held+2019, Bronnikov2021, Mazza+2021, Carballo-Rubio+2022, Eichhorn+2022, Zhou+2023}. Understandably, such matter typically violates the classical energy conditions close to the center of the spacetime \cite{Zaslavskii2010, Bambi+2013}. Attempts to construct regular BH models that avoid the blueshift instability have also been similarly instructive  \cite{Frolov2016, Carballo-Rubio+2022, Casadio+2023}, indicating, e.g., that inner horizons should have vanishing surface gravity.

Non-Kerr BH models can be used to make concrete observable predictions, which can then be confronted with observations, e.g., with the recent images of M87$^*$ and Sgr A$^*$, to examine consistency (see, e.g., \cite{Kocherlakota+2021, EHTC+2022f, Vagnozzi+2023}) and, in principle, direct theoretical efforts. In addition to the immense wealth of analytical work on this particular theme (see, e.g., Refs. \cite{Abdujabbarov+2016, Amir+2016, Cunha+2018, Cardoso+2019, Kumar+2019, Banerjee+2022, Chen2022, Kumar+2022}), numerical simulations of magnetized accretion in non-Kerr spacetimes are now being performed to enable increasingly robust comparisons with the observed images of supermassive BHs \cite{Mizuno+2018, Olivares+2020, Nampalliwar+2022, Roder+2023, Chatterjee+2023a, Chatterjee+2023b, Combi+2024}. Planned ground-based \cite{Doeleman+2023, Johnson+2023} and space-based \cite{Lupsasca+2024, Johnson+2024} upgrades to the current telescope arrays used to obtain such measurements will only tighten existing constraints on non-Kerr spacetimes (see, e.g., Refs. \cite{Johnson+2020, Gralla+2020, Ayzenberg+2023, Kocherlakota+2024}).

General-relativistic non-Kerr BH spacetimes typically have nontrivial Einstein tensors, and, therefore, from the Einstein equation, can be seen to generically contain nontrivial matter content. Due to the increasing importance of non-Kerr BH models, either in theoretical inquiries or in observational tests of gravity, it is equally important to understand the physical properties of the spacetime-generating, self-gravitating matter.

Our primary goal here is to examine how variations in the spacetime geometry affect the physical properties of the associated self-gravitating matter (e.g., its flow profile and whether or not it satisfies the energy conditions), both in the BH exterior as well as in its \textit{interior}. We assume throughout that such matter is minimally-coupled to gravity. Interestingly, we demonstrate here that the existence of a stationary horizon imposes novel constraints on the \textit{types} of matter (as in Sec. 4.3 of Ref. \cite{Hawking+1973}) that are permitted to thread it. 

To enable a systematic investigation of the changes in properties of the self-gravitating matter caused by modifying the spacetime geometry, we consider a broad family of stationary and axisymmetric spacetime metrics that satisfy the Azreg-A\"inou (AA) metric \textit{ansatz} \cite{Newman+1965, Azreg-Ainou2014a, Azreg-Ainou2015}. The AA metric \textit{ansatz} (Sec. \ref{sec:SecII_Metric_Ansatz}) uses two free metric functions to describe asymptotically-flat, spinning and geodesically-integrable (i.e., all geodesics possess a Carter constant) spacetimes. For a specific choice of the metric functions, the AA metric describes the Kerr-Newman BH spacetime \cite{Newman+1965b}, the simplest spinning solution of the Einstein-Maxwell equations. Therefore, our analysis below corresponds to an exploration of an interesting slice of non-vacuum BH spacetimes, in the vicinity of the Kerr metric, in the phase space of GR. Additionally, its simple algebraic structure is perfectly suited for our demonstrative purposes, ensuring both physical clarity as well as computational economy. For example, the associated Einstein tensor has a single independent off-diagonal component in Boyer-Lindquist (BL; \cite{Boyer+1967}) coordinates ($\mathscr{G}_{t\varphi}$), which greatly simplifies identifying the matter rest-frame (Sec. \ref{sec:SecIII_Comoving_Frame}). For similar reasons, the AA metric has been employed to explore the impact of varying the spacetime geometry on, e.g., the observable BH shadow \cite{Abdujabbarov+2016, Shaikh2019, Junior+2020} or on the quasinormal mode spectrum \cite{Jusufi+2021} in \textit{spinning} BH geometries. 

The \textit{interior} geometry of a generic BH spacetime contains non-stationary regions (``interior cosmologies''). For the Kerr metric, in particular, the spacetime between the horizons corresponds to a novel vacuum, \textit{spinning} cosmology. We isolate such regions in AA BH spacetimes (Sec. \ref{sec:SecIIB_BH_Interior}) and introduce a new set of BL-like coordinates to describe them (Sec. \ref{sec:SecIIC_Interior_Coordinates}). Unlike typical coordinate systems used to cover the BH interior, which are adapted to (principal) \textit{null} congruences, these coordinates are adapted to a congruence of \textit{timelike} observers.%
\footnote{However, see also the Doran coordinates \cite{Doran2000, Wiltshire+2009} or the Painlev{\'e}-Gullstrand coordinates \cite{Blau2023} or the spherical-ingoing Kerr-Schild coordinates \cite{Font+1998, McKinney+2004, Johannsen2013, Kocherlakota+2023}).} %
These coordinates reveal the background matter to be comoving with the interior cosmology (Sec. \ref{sec:SecIII_Comoving_Frame}). 

Thus, to understand the properties of the background matter in the BH interior (e.g., its motion), we are compelled to understand better the properties of the spacetime geometry there. The interior coordinates allow us to transparently characterize the interior geometry, revealing, e.g., the highly nontrivial topology of spacelike hypersurfaces (Sec. \ref{sec:SecIIC_Interior_Coordinates}; See also Refs. \cite{Brehme1977, Doran+2006} for a discussion on the Schwarzschild interior). 

In Sec. \ref{sec:SecIID_PNCs}, we obtain simple analytic expressions for the future expansions of the two zero angular momentum null congruences \textit{throughout} the spacetime. In these spacetimes, these null geodesics remain on conical surfaces ($\vartheta=\mathrm{const.}$), either spiralling monotonically inwards to the center $r=0$ or outwards to $r=\infty$, respectively (see also, e.g., Sec. 33.6 of Ref. \cite{Misner+1964}). The null expansions determine the fractional rate of change of the cross-sectional areas of these null congruences (see, e.g., Sec. 2.4.8 of Ref. \cite{Poisson2004}). Thus, these provide new, and more concrete, geometric intuition of the interior cosmology of spinning BHs. 

In the first portion of this work (Sec. \ref{sec:SecIII_Comoving_Frame}), we will remain agnostic about the specific matter fields that generate the AA metric. The resolution between different types of matter will be coarse, i.e., we will only be able to distinguish them at the level of the energy-momentum-stress tensor. As we emphasize below, this, nevertheless, allows us to build important general intuition regarding their physical properties in BH spacetimes. 

We develop a framework that allows us to use the Einstein equation to identify the legitimate comoving-frame or rest-frame of fluid matter as well as of (e.g., electromagnetic) \textit{fields} (see the quadratic equation \ref{eq:Omega_Quadratic}). We will use this to describe, e.g., the angular velocity profile of the electromagnetic field in the Kerr-Newman spacetime in Sec. \ref{sec:SecV_Spinning_BH_Properties}. While the matter $4-$ velocity is shown to be given by a general Killing vector at each point in the stationary regions of these AA spacetimes, we uncover an additional constraint (i.e., in addition to the classical energy conditions) on the physical viability of an arbitrary AA spacetime geometry. If the matter $4-$velocity is determined to be spacelike for specific choices of the metric functions then such choices can be rejected as being unphysical. Furthermore, we also find that generic choices for the metric functions can lead to the matter exhibiting non-rigid rotation on BL coordinate $2-$spheres (in addition to the differential rotation across $2-$spheres) in these \textit{geodesically-integrable} spacetimes. In our view, this is a non-trivial update to preceding work that has typically \textit{demanded} that the matter be rigidly-rotating on each sphere, and at a particular angular velocity (e.g., Refs. \cite{Azreg-Ainou2014a, Azreg-Ainou2014b, Azreg-Ainou2014c}). More precisely, the matter rest frame has previously been assumed to be given by the Carter tetrad \cite{Carter1968, Znajek1977}.

As mentioned above, we also obtain nontrivial constraints on the types of matter that are permitted to thread stationary event horizons. Given a unique matter rest-frame, it is straightforward to unambiguously introduce the rest-frame energy density, $\epsilon$, and the three principal pressures, $p_i$. In Sec. \ref{sec:SecIIIA_Horizons}, we demonstrate that matter on the horizon, including in spinning spacetimes, always satisfies 
\begin{equation}
p_n = -\epsilon\,,    
\end{equation}
where $p_n$ is the principal pressure (or tension) normal to spherical surfaces. We point out that this result can be used to set up useful ``no-go'' theorems for BH solutions. An example of the latter is that static and spherically-symmetric BHs cannot be produced by a massless (real) scalar field that is minimally-coupled to gravity (\ref{app:AppF1_Scalar_Fields}; cf. also Ref. \cite{Bekenstein1972}). We expect this result to have interesting implications for the Penrose weak cosmic censorship conjecture \cite{Penrose2002}. In Sec. \ref{sec:SecIIIB_Degenerate_Spacetimes}, we discuss a special class of ``degenerate'' nonspinning spacetimes as well as their spinning counterparts, whose matter content satisfies $p_n = -\epsilon$ everywhere. Electromagnetic fields are an example of such matter (\ref{app:AppF2_EM_Fields}). 

We emphasize that the analysis framework we develop below should continue to apply for more general classes of stationary and axisymmetric spacetimes (see, e.g., Refs. \cite{Vigeland+2011, Gair+2011, Johannsen2013, Rezzolla+2014, Konoplya+2016, Delaporte+2022}). Although not the focus of this work, judging the physical reasonability of the background matter (e.g., via the classical energy conditions) can also introduce theoretical priors on permissible deformations of the Kerr metric in various frameworks used when performing parametric tests of gravity \cite{Psaltis+2020, EHTC+2022f}.

In the second portion of this work (Secs. \ref{sec:SecIV_Nonspinning_BH_Properties}, \ref{sec:SecV_Spinning_BH_Properties}), we discuss specific GR solutions, whose matter content is known, in order to make contact with, and verify, the general findings from previous sections.

Applying the framework developed in Sec. \ref{sec:SecIII_Comoving_Frame} to analyze the physical properties of self-gravitating matter in the AA metric, in Sec. \ref{sec:SecIV_Nonspinning_BH_Properties} we present the properties of matter -- the matter flow profile as well as the rest-frame energy density and principal pressure distribution -- in a variety of popular \textit{non-spinning} BH spacetimes containing different kinds of matter fields. These include the Reissner-Nordstr{\"o}m (RN) BH solution (the non-spinning electromagnetically charged BH of GR), the Gibbons-Maeda-Garfinkle-Horowitz-Strominger (GMGHS) BH solution \cite{Gibbons+1988, Garfinkle+1991} (the counterpart of the RN BH in the low-energy effective theory of the heterotic string), and several phenomenologically-proposed regular BH models that contain anisotropic fluid matter \cite{Hayward2006, Frolov2016, Zhou+2023}. We note that the GMGHS BH can equivalently be seen as a solution of GR minimally-coupled to scalar and electromagnetic fields, and can thus be treated within our framework. We recover some well-known results and also report new results. 

In Sec. \ref{sec:SecV_Spinning_BH_Properties}, we analyze the self-gravitating matter in the spinning counterparts of the metrics in Sec. \ref{sec:SecIV_Nonspinning_BH_Properties}, all of which are described by the AA metric \textit{ansatz}, to understand the impact of a smoothly varying spin on matter and geometry. These will include the Kerr-Newman \cite{Newman+1965b} and the Kerr-Sen \cite{Sen1992} solutions, which are the legitimate stationary generalizations of the RN and GMGHS solutions respectively, for a non-vanishing spin parameter. We comment on the extent to which the classical energy conditions are satisfied by the underlying matter distributions in these spinning spacetimes, and also explore the rotation characteristics of the matter, specifically the degree of differential rotation present.

Finally, we summarize our conclusions in Sec. \ref{sec:SecVI_Conclusions}.

Throughout we adopt geometrized units, in which $G=c=1$, and the metric signature is $(-, +, +, +)$. We reserve $M$ to denote the total Arnowitt-Deser-Misner (ADM; \cite{Arnowitt+2008}) mass of the spacetime. Bracketed indices indicate that the corresponding tensor is projected onto a tetrad. Tensor components in static spacetimes are demarcated by hats. 


\section{Axisymmetric Metric Ansatz}
\label{sec:SecII_Metric_Ansatz}

The line-element of a general static and spherically-symmetric metric, $\hat{\mathscr{g}}_{\mu\nu}$, can be written in spherical-polar coordinates, $\hat{x}^\mu = (t, r, \vartheta, \varphi)$, as
\begin{equation} \label{eq:Spherically_Symmetric_Metric}
\mathrm{d}s^2 = -f\mathrm{d}t^2 + \frac{g}{f}~\mathrm{d}r^2 + R^2~\mathrm{d}\Omega_2^2\,,
\end{equation}
where $\mathrm{d}\Omega_2^2 = \mathrm{d}\vartheta^2 + \sin^2{\vartheta}~\mathrm{d}\varphi^2$ is the standard line-element on a unit $2-$sphere, and the metric functions $f, g,$ and $R$ are functions of $r$ alone. For the Schwarzschild metric, in particular, $f(r) = 1-2M/r, g(r) = 1$, and $R(r) = r$. 

The metric function $R(r)$ measures the areal or curvature radius of a $2-$sphere of coordinate radius $r$. It increases monotonically with $r$, and is required to be nonnegative for $r>r_0$ and permitted to (but not required to; cf. Refs. \cite{Kazakov+1993, Modesto2008}) vanish at $r=r_0$.%
\footnote{For most of the spacetimes considered here $r_0 = 0$. The only exception is the GMGHS spacetime (Table \ref{table:Known_Static_Solutions}), for which $r_0 = Q^2/M$. Note, however, that by a simple change of coordinates, we can achieve $r_0 = 0$ for the GMGHS metric (see, e.g., Ref. \cite{Kocherlakota+2020}).} %
The metric function $g(r)$ is required to be positive-definite to preserve the Lorentzian nature of the metric everywhere: The determinant of the metric is given by $\mathrm{det}[\hat{\mathscr{g}}_{\mu\nu}] = -g R^4\sin^2{\vartheta}$. When horizons exist, they are located at the zeroes of $\hat{\mathscr{g}}^{rr}$, i.e., at the roots of $f(r)$. 

We can also trade the metric functions $f$ and $g$ for two new metric functions $m$ and $\psi$ so that the general spherically-symmetric metric \eqref{eq:Spherically_Symmetric_Metric} takes the form
\begin{equation} \label{eq:Static_Metric}
\begin{aligned}
\mathrm{d}s^2 
=&\ -\mathrm{e}^{2\psi}\left(1-\frac{2m}{R}\right)\mathrm{d}t^2 + \frac{(\partial_r R)^2}{\left(1-\frac{2m}{R}\right)}~\mathrm{d}r^2 + R^2~\mathrm{d}\Omega_2^2\,.
\end{aligned}
\end{equation}
The metric function $m(r)$ now corresponds to the Hawking mass \cite{Hawking1968} or the Misner-Sharp-Hernandez mass \cite{Misner+1964, Hernandez+1966} of the spacetime and $\psi(r)$, the redshift function, is related to the matter pressure distribution. For the Schwarzschild metric, in particular, we have $R(r) = r, m(r) = M$, and $\psi(r) = 0$. The metric functions $f$ and $g$ are related to $m$ and $\psi$ clearly as 
\begin{equation} \label{eq:fg_m_psi}
\begin{aligned}
f(r) = \mathrm{e}^{2\psi}\left(1 - \frac{2m}{R}\right)\,;\ \ 
g(r) = \left(\mathrm{e}^{\psi}\cdot\partial_rR\right)^2\,.
\end{aligned}
\end{equation}
We will use both versions (\ref{eq:Spherically_Symmetric_Metric}, \ref{eq:Static_Metric}) interchangeably. 

The Azreg-A{\"i}nou (AA) metric \textit{ansatz}, $\mathscr{g}_{\mu\nu}$, which describes a stationary and axisymmetric spacetime, is given, in Boyer-Lindquist (BL; \cite{Boyer+1967}) coordinates, $x^{\mu} = (t, \rho, \vartheta, \varphi)$, as \cite{Azreg-Ainou2015, Chen2022a} 
\begin{equation} \label{eq:Stationary_Metric}
\begin{aligned}
\mathrm{d}s^2
=&\ 
-\left(1-\frac{2 F}{\Sigma}\right)\mathrm{d}t^2 
-2\frac{2 F}{\Sigma}a\sin^2{\vartheta}~\mathrm{d}t\mathrm{d}\varphi \\ 
&\ 
+ \frac{\Pi}{\Sigma}\sin^2{\vartheta}~\mathrm{d}\varphi^2 + \frac{\Sigma}{\Delta}g~\mathrm{d}r^2 + \Sigma~\mathrm{d}\vartheta^2\,.
\end{aligned}
\end{equation}
In the above, $a$ denotes the specific angular momentum of the spacetime. In terms of two auxiliary functions $A$ and $B$, which are related to the static and spherically-symmetric ``seed'' metric functions ($f, R$) in eq. \ref{eq:Spherically_Symmetric_Metric} as
\begin{equation} \label{eq:Auxiliary_Metric_Functions}
\begin{aligned}
A(r) =\ R^2(r)\,;\ 
B(r) =\ f(r) R^2(r)\,,
\end{aligned}
\end{equation}
the AA metric functions are given as,
\begin{equation} \label{eq:Stationary_Metric_Functions}
\begin{alignedat}{3}
&\ F = (A - B)/2\,;\ 
&&\ \Delta = B + a^2\,, \\
&\ \Sigma = A + a^2\cos^2{\vartheta}\,;\ 
&&\ \Pi = \left(A + a^2\right)^2 - \Delta a^2\sin^2{\vartheta}\,.  
\end{alignedat}
\end{equation}
As can be seen from the above, $F$ and $\Delta$ are functions of $r$ alone, whereas $\Sigma$ and $\Pi$ are functions of both $r$ and $\vartheta$.

Notice that the zero-spin limit ($a=0$) of the spinning metric \eqref{eq:Stationary_Metric} reduces to the nonspinning metric \eqref{eq:Spherically_Symmetric_Metric} exactly. We will use the metric \eqref{eq:Stationary_Metric} as the \textit{ansatz} for the spacetime geometry in this work.

This choice for the metric \textit{ansatz} has the attractive property that if the metric \eqref{eq:Spherically_Symmetric_Metric} describes an asymptotically-flat spacetime, the metric \eqref{eq:Stationary_Metric} is also assured to be asymptotically-flat \cite{Kocherlakota+2023}. Furthermore, this class of spacetimes is geodesically-integrable \cite{Azreg-Ainou2014c, Kocherlakota+2024}, i.e., all geodesics possess a Carter constant \cite{Carter1968}. 

It is also worth noting that the AA metric was originally conceived through a modification of the Newman-Janis (NJ) solution-generating algorithm \cite{Newman+1965}. Due to the complexity involved in obtaining stationary solutions to the coupled Einstein-matter field equations, solution-generating techniques are often employed. For example, given a generic static and spherically-symmetric ``seed'' metric that solves appropriate field equations (Einstein-Klein-Gordon, Einstein-Maxwell, etc.), one attempts to algorithmically produce, by leveraging mathematical tools such as coordinate transformations, a legitimate stationary and axisymmetric counterpart, without having to resolve the field equations. 

The NJ-AA algorithm is currently the most popular algorithm (but see also Ref. \cite{Contreras+2021}), and has been very useful in producing reasonable spacetime geometries that are ostensibly spinning generalizations of static regular BHs that are sourced by anisotropic fluids \cite{Azreg-Ainou2014c} or by nonlinear electrodynamics fields \cite{Toshmatov+2017}, from loop quantum gravity \cite{Liu+2020}, Yang-Mills theory \cite{Jusufi+2021}, of wormholes \cite{Azreg-Ainou2014b}, and of rotating BH mimickers in general \cite{Mazza+2021}. Nevertheless, it is important to note that the NJ-AA algorithm is a ``partial'' solution-generating algorithm. Its nontrivial success is in producing an analytic stationary metric \textit{ansatz} -- with the attractive algebraic qualities mentioned above -- that reduces to the seed metric smoothly in the limit of vanishing spin. This simple \textit{ansatz} metric can then be used to cleanly infer the matter energy-momentum-stress tensor from the gravitational field equations (thus solving them by construction), which enables a clear investigation of its physical reasonability (as in Sec. \ref{sec:SecIII_Comoving_Frame}). However, obtaining explicit solutions to the \textit{matter} field equations (for the associated matter field profiles) in a similar fashion remains challenging in general (see, however, Ref. \cite{Erbin2017}). We direct the reader to \ref{app:AppA_AA_Framework} for further discussion.


\subsection{Properties of the Ansatz Metric}
\label{sec:SecIIA_Ansatz_Metric_Properties}

Ergosurfaces or static limit surfaces (see, e.g., Ref. \cite{Wiltshire+2009}) of the spacetime \eqref{eq:Stationary_Metric} are defined as locations where the asymptotically timelike Killing vector $\mathbf{T} = \partial_t$ changes character, i.e., where $\mathscr{g}_{tt} = 0.$:
\begin{equation} \label{eq:Ergosurfaces}
{\rm Ergosurfaces:} \quad B + a^2\cos^2{\vartheta} = 0\,.
\end{equation}

When the metric \eqref{eq:Stationary_Metric} is used to describe black hole (BH) spacetimes, its event horizon is a stationary, null hypersurface \cite{Hawking1972}, and is generated by the null Killing vector,
\begin{equation} \label{eq:Horizon_Generator}
\ell_{\mathrm{H}}^\mu = \left[1, 0, 0, \Omega_{\mathrm{H}}\right]\,;\ \ \Omega_{\mathrm{H}} = \frac{a}{A(r_{\mathrm{H}})+a^2} = \frac{a}{\mathscr{R}_{\mathscr{A}; \mathrm{H}}^2}\,.
\end{equation}
Here $\Omega_{\mathrm{H}}$ denotes the angular velocity of the event horizon. Equivalently, since the horizon is a constant$-r$ null hypersurface, its normal, $\nabla_\mu r$, satisfies $\mathscr{g}^{\mu\nu}(\partial_\mu r)(\partial_\nu r) = \mathscr{g}^{rr} = 0$. Its location, $r=r_{\mathrm{H}}$, therefore, corresponds to the largest positive root of $\Delta$. Inner horizons correspond to other roots of $\Delta$, hence
\begin{equation} \label{eq:Horizons}
{\rm Horizons:}\quad \Delta = B + a^2 = 0\,.
\end{equation}
In the above \eqref{eq:Horizon_Generator}, $\mathscr{R}_{\mathscr{A}}$ denotes the areal radius of a BL coordinate $2-$sphere of radius $r$, i.e., 
\begin{equation} \label{eq:Areal_Radius}
\mathscr{R}_{\mathscr{A}}(r) = \sqrt{\mathscr{A}(r)/4\pi}\,,
\end{equation}
where $\mathscr{A}$ is its area. \ref{app:AppB_CTCs} reports an analytic expression for the latter in this broad class of stationary spacetimes. The relation of the horizon areal radius, $\mathscr{R}_{\mathscr{A}; \mathrm{H}}$, to its angular frequency sets up a loose comparison between the event horizon and a uniform spherical \textit{solid} that is rigidly-rotating about its axis in Newtonian mechanics. The angular momentum, $J$, mass, $M$, angular frequency, $\Omega$, and radius, $\mathscr{R}$, of the latter are related as $J/M = (2/5)\mathscr{R}^2\Omega$. For a rigidly-rotating spherical \textit{surface} instead, we have $J/M = (2/3)\mathscr{R}^2\Omega$.

Asymptotic flatness demands that $\lim_{r\rightarrow\infty}\mathscr{g}_{tt} = -1$, i.e., $\lim_{r\rightarrow\infty}[-(B+a^2\cos^2{\vartheta})/(R^2+a^2\cos^2{\vartheta})] = -1$. This implies that $\lim_{r\rightarrow\infty}B \approx R^2$. Thus, since $B > 0$ at asymptotic infinity, as we move closer to the BH, condition \eqref{eq:Ergosurfaces} will be met before the condition \eqref{eq:Horizons}. This ensures that the outermost ergosurface always lies outside the event horizon. The region bounded by the two surfaces is referred to as the (exterior) ergoregion. For nonspinning BH spacetimes ($a=0$), the event horizon is a static ($\Omega_{\mathrm{H}} = 0$) null hypersurface and an ergoregion is absent.

The metric \eqref{eq:Stationary_Metric} has a coordinate singularity at the location of the event horizon: $\Delta = 0 \Rightarrow \mathscr{g}_{rr} \rightarrow \infty$. However, if we assume that the metric components are real, analytic functions, it is possible to ``extend'' the spacetime by moving to better coordinates and adding new regions in the standard way \cite{Eddington1924, Finkelstein1958, Rindler1960, Kruskal1960, Szekeres2002, Penrose1962, Kerr+2009} (see also \ref{app:AppC_Null_Kerr_Schild} below). 

Before proceeding with extending the spacetime past the coordinate singularity in Sec. \ref{sec:SecIIB_BH_Interior}, we conclude this section by describing the principal null congruences (PNCs) of this spacetime. Members of these congruences (the principal null geodesics) along with the horizon null generators encode the global causal structure of the spacetime. The outgoing ($+$) and ingoing ($-$) PNCs, generated by the vector fields $\ell_\pm$, are obtained as solutions to (see, e.g., eq. 4.16 of Ref. \cite{Stephani+2003})
\begin{equation}
\ell^\mu\ell_{[\alpha}\mathscr{C}_{\beta]\mu\nu[\gamma}\ell_{\delta]}\ell^\nu = 0\,,
\end{equation}
where $\mathscr{C}_{\beta\mu\nu\gamma}$ is the Weyl tensor. In BL coordinates, for these spacetimes, these are given as (see also Ref. \cite{Kocherlakota+2023})
\begin{equation} \label{eq:PNC_Vector_Fields_BL}
\ell_\pm^\mu = \left[(A + a^2)\frac{\sqrt{g}}{\Delta}, \pm 1, 0, a\frac{\sqrt{g}}{\Delta}\right]\,.
\end{equation}
As can be seen from the above, the integral curves generated by these vector fields remain on conical surfaces (since $\ell^\vartheta_\pm$ is identically zero) and have no radial turning points, i.e., they either monotonically approach the center ($r=r_0; -$) or infinity ($r=\infty; +$). At a horizon, these limit to its null generators, as given in eq. \ref{eq:Horizon_Generator}. The description above matches that in Sec. 33.6 of Ref. \cite{Misner+1973}, for the case of the Kerr metric.

As written in eq. \ref{eq:PNC_Vector_Fields_BL}, the vector fields generate non-affinely parametrized principal null geodesics, i.e., they satisfy the geodesic equation of the form $\ell^\mu\nabla_\mu\ell^\nu = \kappa\ell^\nu$ with $\kappa = \partial_rg/(2g)$. Dividing both vector fields in eq. \ref{eq:PNC_Vector_Fields_BL} by $\sqrt{g}$ leads to affinely parametrized principal null geodesics (i.e., $\kappa = 0$).


\subsection{The Black Hole Interior}
\label{sec:SecIIB_BH_Interior}

We note that thus far we have referred to the metrics in eqs. \ref{eq:Static_Metric} and \ref{eq:Stationary_Metric} as being static and stationary, respectively. While this is true in the region of spacetime outside the event horizon (``region I''; See, e.g., Ch. 5 of Ref. \cite{Poisson2004}), it is not true for the maximally-extended spacetime. 

This is seen by considering the character of the general Killing vector, $\mathbf{K}$. With $\mathbf{T}=\partial_t$ the (asymptotically-timelike) time-translation Killing vector and $\mathbf{\Phi} = \partial_\varphi$ the (asymptotically-spacelike) rotational Killing vector, the norm of the general Killing vector, $\mathbf{K} = \mathbf{T} + \Omega \mathbf{\Phi}\ (\Omega=\mathrm{const.}$),
\begin{equation} \label{eq:Killing_Norm}
N(\Omega) := \mathscr{g}_{\mu\nu}K^\mu K^\nu = \mathscr{g}_{tt} + 2\Omega \mathscr{g}_{t\varphi} + \Omega^2\mathscr{g}_{\varphi\varphi}\,,
\end{equation}
is negative (timelike) in the exterior horizon geometry if $\Omega_- < \Omega < \Omega_+$, and is positive (spacelike) otherwise. Here
\begin{equation} \label{eq:Null_Angular_Velocities}
\Omega_{\pm} = \frac{-\mathscr{g}_{t\varphi} \pm \sqrt{-\Delta_{t\varphi}}}{\mathscr{g}_{\varphi\varphi}} = \frac{2aF \pm \sqrt{\Delta}\Sigma\csc{\vartheta}}{\Pi}
\end{equation}
are solutions to $N(\Omega) = 0$, and $\Delta_{t\varphi} := \mathscr{g}_{tt}\mathscr{g}_{\varphi\varphi} - \mathscr{g}_{t\varphi}^2 = -\Delta\sin^2{\vartheta}$ denotes the determinant of the $t\varphi-$sector of the metric. Thus, the Killing vectors $\mathbf{K_\pm} = \mathbf{T} + \Omega_\pm\mathbf{\Phi}$ are null. For completeness, we note that the angular velocity of the zero angular momentum observer (ZAMO), $\Omega_{\mathrm{Z}}$, is given as 
\begin{equation}
\Omega_{\mathrm{Z}} = -\frac{\mathscr{g}_{t\varphi}}{\mathscr{g}_{\varphi\varphi}} = \frac{2aF}{\Pi}\,.   
\end{equation}
At the event horizon, the critical null angular velocities $\Omega_\pm$ become $\vartheta-$independent since $\Delta = 0$, and we find $\Omega_+(r_{\mathrm{H}}) = \Omega_-(r_{\mathrm{H}}) = \Omega_{\mathrm{Z}}(r_{\mathrm{H}}) = \Omega_{\mathrm{H}}.$


\begin{figure}
\centering
\includegraphics[width=\columnwidth]{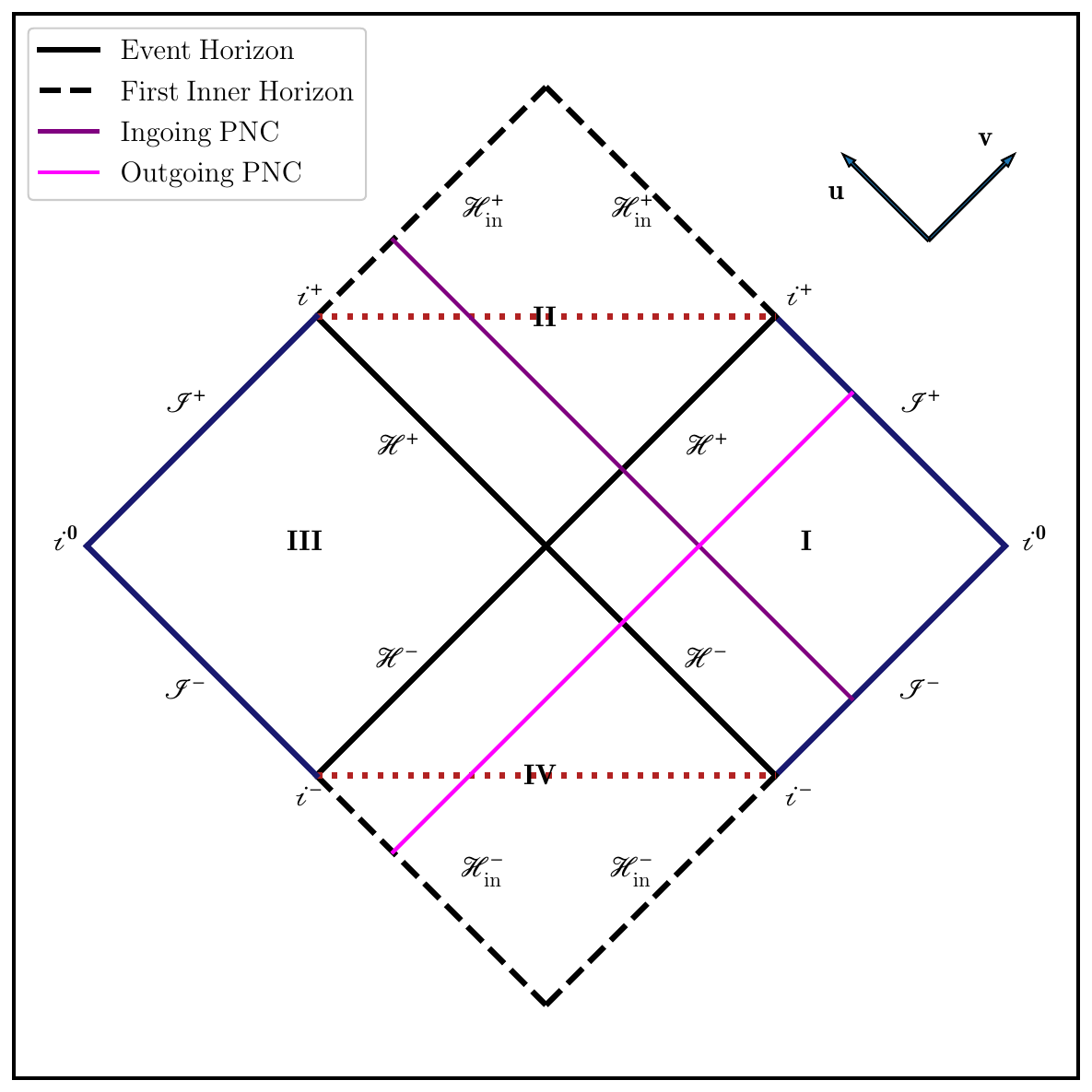}
\caption{Shown here are four regions (I-IV) of the Penrose diagram of a typical asymptotically-flat and -stationary nonextremal black hole spacetime. The orientation of the $u$ and $v$ axes, shown in the top right corner, makes the region numbering clearer. Here, $\mathscr{I}^\pm$ denote the future ($+$) and the past ($-$) null infinities, $\mathscr{i}^\pm$ denote the future ($+$) and the past ($-$) timelike infinities, and $\mathscr{i}^0$ denotes spacelike infinity. The future ($+$) and past ($-$) event ($\mathscr{H}$) and first inner ($\mathscr{H}_{\mathrm{in}}$) horizons are also marked. As shown, members of the ingoing ($-$) principal null congruence (PNC) correspond to constant$-v$ null geodesics \eqref{eq:PNC_inKS}. Similarly, members of the outgoing ($+$) PNC correspond to constant$-u$ lines in this diagram. While Boyer-Lindquist coordinates \eqref{eq:Stationary_Metric} can be used to conveniently describe the spacetime, these are valid only in region I, due to coordinate singularities at the event horizon. Ingoing null (Kerr; \cite{Kerr1963}) coordinates allow us to extend the spacetime through to region II (\ref{app:AppC_Null_Kerr_Schild}). Using the latter, we can then introduce a new set of Boyer-Lindquist-like coordinates to describe region II of the black hole interior (Sec. \ref{sec:SecIIC_Interior_Coordinates}). To obtain the maximally-extended spacetime, many such extension regions are necessary in general. Finally, for a static black hole spacetime with a single horizon and a spacetime singularity (e.g., Schwarzschild), the Penrose diagram terminates at the red dotted lines.}
\label{fig:Fig1_Penrose_Diagram}
\end{figure}


Immediately inside the (future) event horizon (see, e.g., region II of Fig. \ref{fig:Fig1_Penrose_Diagram}), however, $\Omega_\pm$ cease to be real since  $\Delta < 0$. Thus, $N(\Omega)$ must have the same sign for all $\Omega$, and is positive, in particular, since $N(\Omega=0) = (a^2\sin^2{\vartheta} - \Delta)/\Sigma >0$. Therefore, while there exist Killing vectors that are timelike ($N<0$) outside the event horizon, none remain timelike just inside it. This is to be expected since the event horizon is also a Killing horizon. 

Fig. \ref{fig:Fig1_Penrose_Diagram} shows four regions of the Penrose diagram \cite{Penrose1962} of a typical nonextremal BH spacetime, obtained by extending the metric in eq. \ref{eq:Stationary_Metric}. The set of BL coordinates, $x^\mu = (t, r, \vartheta, \varphi)$, in \eqref{eq:Stationary_Metric} cover region I, i.e., the principal region, corresponding to the exterior horizon geometry. We designate the region between the future event horizon, $\mathscr{H}^+$, and the first future inner horizon, $\mathscr{H}_{\mathrm{in}}^+$, (this is located at the second largest positive root of $\Delta = 0$) as region II. Similarly, we designate the region between the past event horizon, $\mathscr{H}^-$, and the first past inner horizon, $\mathscr{H}_{\mathrm{in}}^-$, as region IV. The first inner horizon typically corresponds to a Cauchy horizon. Region III describes the exterior horizon geometry in a different asymptotic universe. The maximally-extended spacetime contains infinite such regions in general. For spacetimes containing a single horizon and a curvature singularity, such as the Schwarzschild BH spacetime, the Penrose diagram truncates at the red dotted lines. 

The absence of a timelike Killing vector in several regions (e.g., II, IV) of the BH interior implies that the spacetime is not stationary there. Instead, it corresponds to a cosmological spacetime (see, e.g., Refs. \cite{Brehme1977, Doran+2006} for a discussion on the Schwarzschild BH interior as well as Refs. \cite{Bianchi2001, Kantowski+1966}). This is a generic feature of the maximally-extended spacetime: regions in which $\Delta > 0$ are stationary and those in which $\Delta < 0$ are not. These are separated by the horizons ($\Delta=0$).

The positivity of $g(r)$ for $r \geq r_0$ implies that the determinant of the spacetime metric $\mathrm{det}[\mathscr{g}_{\mu\nu}] = -\Sigma^2 g \sin^2{\vartheta}$ is assuredly negative-definite, and, thus, that the spacetime \eqref{eq:Stationary_Metric} is Lorentzian over $r\geq r_0$.

The denominators of the Ricci and Kretschmann scalars are proportional to $\propto \Sigma^3 g^2$ and $\propto \Sigma^6 g^4$, respectively, whereas their numerators are functions of the metric functions $g, A, B$, and their first and second derivatives. Assuming these metric functions are twice-differentiable, if the spacetime contains a curvature singularity, it must occur at $\Sigma = 0$. In spinning singular spacetimes, the singularity is located at $r=r_0$ in the equatorial ($\vartheta=\pi/2$) plane. This is because the metric function $\Sigma=R^2(r)+a^2\cos^2{\vartheta}$ inherits its properties from the seed metric function $R(r)$, which is positive definite on $r > r_0$ and is permitted to vanish at $r=r_0$. 

Henceforth, including when referring to the maximally-extended spacetime, we will restrict our attention to regions of spacetime in which there are no closed timelike curves (CTCs; \cite{Thorne1992}), i.e., we focus on regions in which $\mathscr{g}_{\varphi\varphi} > 0$ (see Footnote \ref{fn:CTCs}), or equivalently, $\Pi > 0$. Although possible to do, we do not extend the metric interior to the outermost surface $r=r_{\mathrm{CTC}}$ where $\Pi(r_{\mathrm{CTC}}, \vartheta)=0$. This is discussed further in \ref{app:AppB_CTCs} and Sec. \ref{sec:SecVA_CTCs}.

The statements above relating to the regions obtained by analytically extending the spacetime beyond the original region can be made precise by using better (e.g., Rindler \cite{Rindler1960}, ingoing null \cite{Eddington1924, Finkelstein1958, Kerr1963, Kerr+2009}, or double null \cite{Kruskal1960, Szekeres2002, Penrose1962}) coordinates that actually cover the interior. 

In \ref{app:AppC_Null_Kerr_Schild}, by performing a coordinate transformation, we introduce coordinates adapted to a congruence of radially-ingoing photons. These ingoing null (Kerr; \cite{Kerr1963}) coordinates allow us to extend the spacetime smoothly across the future event horizon. In addition to providing a clear interpretation of the character of the asymptotically-timelike Killing vector $\mathbf{T}$ as well as the general Killing vector $\mathbf{K}$, these coordinates also expose a crucial geometrical interpretation of the coordinate $r$ as the affine parameter along such geodesics, throughout the maximally-extended spacetime.


\subsection{Interior Coordinates}
\label{sec:SecIIC_Interior_Coordinates}

In this section, we introduce a set of Boyer-Lindquist-like coordinates, $x^{\breve{\mu}} = (\tau, z, \vartheta, \phi)$, for the BH \textit{interior}. These coordinates allow us to define a useful notion of ``cosmological time'' for the interior \textit{spinning} geometry. These will be valid specifically in region II of the maximally-extended spacetime, and are obtained from the aforementioned ingoing null coordinates via the following coordinate transformation (effectively the inverse of equation \ref{ingoingcoords}),
\begin{equation}
\tau = r\,,\ z = v - r_\star\,,\ \phi = \bar{\phi} - r_\blacklozenge\,.
\end{equation}
The line element in these coordinates becomes,
\begin{align} \label{eq:Stationary_Metric_iBL}
\mathrm{d}s^2
=&\ 
\frac{\Sigma}{\Delta}g~\mathrm{d}\tau^2  + \Sigma~\mathrm{d}\vartheta^2 -\left(1-\frac{2 F}{\Sigma}\right)\mathrm{d}z^2 \\
&\
-2\frac{2 F}{\Sigma}a\sin^2{\vartheta}~\mathrm{d}z\mathrm{d}\phi
+\frac{\Pi}{\Sigma}\sin^2{\vartheta}~\mathrm{d}\phi^2\,. \nonumber 
\end{align}

We see immediately that $\partial_\tau$ is timelike in this region since $\mathscr{g}_{\breve{\mu}\breve{\nu}}(\partial_\tau)^{\breve{\mu}}(\partial_\tau)^{\breve{\nu}} = \mathscr{g}_{\breve{\mu}\breve{\nu}}\delta^{\breve{\mu}}_\tau\delta^{\breve{\nu}}_\tau = \Sigma g/\Delta$, and $\Delta< 0$. We can similarly see that $\partial_\vartheta$ is spacelike. Furthermore, the two Killing vectors are given, in region II, as $\mathbf{T}= \partial_z$ and $\mathbf{\Phi} = \partial_\phi$. The first, $\mathbf{T}$, is spacelike since $\mathscr{g}_{zz} = (2F-\Sigma)/\Sigma$, and $2F > \Sigma$ in this region.%
\footnote{Since $2F = A - B$ and $\Sigma = A + a^2\cos^2{\vartheta}$, this inequality is $B + a^2\cos^2{\vartheta} < 0$, which is true in such regions, where $\Delta = B+a^2 < 0$.} %
We reiterate that here our attention is restricted to the regions of spacetime in which no causality-violation occurs, where $\mathscr{g}_{\varphi\varphi} > 0$.%
\footnote{Since $\partial_t$ and $\partial_\tau$ are timelike in the stationary and nonstationary regions of spacetime respectively, regions for which additionally $\mathscr{g}_{\varphi\varphi} < 0$ are regions in which $\partial_\varphi$ is \textit{also} timelike. Since the integral orbits of $\partial_\varphi$ are closed, these correspond to the regions that permit closed timelike curves, which we ignore in this work. See also the related discussion in \ref{app:AppB_CTCs}.
\label{fn:CTCs}} %
Thus, the general Killing vector $\mathbf{K} = \partial_z + \Omega \partial_\phi\ (\Omega=\mathrm{const.})$ is everywhere spacelike in this region, consistent with the earlier discussion.

Furthermore, the inverse metric is given as,
\begin{equation}
g^{\breve{\mu}\breve{\nu}} = 
\begin{bmatrix}
\frac{\Delta}{\Sigma}\frac{1}{g} & 0 & 0 & 0 \\
0 & -\frac{\Pi}{\Sigma\Delta} & 0 & -\frac{2aF}{\Sigma\Delta}\\
0 & 0 & \frac{1}{\Sigma} & 0 \\
0 & -\frac{2aF}{\Sigma\Delta} & 0 & \frac{-(2F-\Sigma)}{\Sigma\Delta\sin^2{\vartheta}}\\
\end{bmatrix}\,,
\end{equation}
from which we see that the normal to constant$-\tau$ hypersurfaces, $\nabla_{\breve{\mu}}\tau$, is timelike: $\mathscr{g}^{\breve{\mu}\breve{\nu}}(\partial_{\breve{\mu}}\tau)(\partial_{\breve{\mu}}\tau) = \mathscr{g}^{\tau\tau} = \Delta/(\Sigma g)$. Therefore, these are spacelike hypersurfaces. Similarly, the one-forms $\nabla_{\breve{\mu}}z$, $\nabla_{\breve{\mu}}\vartheta$, and $\nabla_{\breve{\mu}}\phi$ are all seen to be spacelike.

We can express the above equivalently in terms of the metric functions of the spacetime in \textit{region I} as,
\begin{align} \label{eq:Interior_Metric_RI}
\mathrm{d}s^2
=&\ \mathscr{g}_{rr}\mathrm{d}\tau^2 + \mathscr{g}_{\vartheta\vartheta}\mathrm{d}\vartheta^2 + \mathscr{g}_{tt}\mathrm{d}z^2 + 2\mathscr{g}_{t\varphi}\mathrm{d}z\mathrm{d}\phi + \mathscr{g}_{\varphi\varphi}\mathrm{d}\phi^2\,, 
\end{align}
to see that, unsurprisingly, the metric in region II \eqref{eq:Stationary_Metric_iBL} is simply the metric in region I \eqref{eq:Stationary_Metric} with the time and ``radial'' coordinates legitimately exchanging character, $r \leftrightarrow \tau$ and $t \leftrightarrow z$. The interior angular coordinates $\phi$ and $\vartheta$ match the exterior ones $\varphi$ and $\vartheta$ precisely, by construction.

Note that the topology of spacelike-hypersurfaces in the nonstationary region II of these spacetimes is radically different from that in the stationary region I (in these coordinates). In particular, $z$ is \textit{not} a radial coordinate because it takes values in $(-\infty, +\infty)$  (cf. Refs. \cite{Brehme1977, Doran+2006}). Nevertheless, its level sets in the hypersurfaces are still spheres. On the other hand, $\tau$ takes values from $r_{\mathrm{H}}$ to $r_{\mathrm{IH}}$, \textit{decreasing} in the future, with $r=r_{\mathrm{IH}} < r_{\mathrm{H}}$ the location of the first inner horizon. We understand this by remembering that $-r$ parameterizes the ingoing PNC, as discussed below eq. \ref{eq:PNC_inKS}.%
\footnote{In region IV, the timelike coordinate, say $\bar{\tau}$, for the interior metric must increase from $r_{\mathrm{IH}}$ to $r_{\mathrm{H}}$ so that $\bar{\tau}$ increases to the future. Again, this can be understood by remembering that $+r$ parameterized the outgoing PNC (see Footnote \ref{fn:Outgoing_Null_Coords}).}

The above is to be kept in mind when interpreting the generators of the PNCs in these coordinates,
\begin{equation} \label{eq:PNC_Vector_Fields_iBL}
\ell_\pm^{\breve{\mu}} = \left[-1, \mp (A + a^2)\frac{\sqrt{g}}{\Delta}, 0, \mp a\frac{\sqrt{g}}{\Delta}\right]\,.
\end{equation}
With this useful coordinate system for the interior, that is adapted to a congruence of \textit{timelike} observers, in hand, we will consider below the properties of matter both in the BH exterior as well as in its interior but only work with coordinate-invariant quantities. The minimal number of off-diagonal entries in the metric tensor, in these coordinates, buys us significant computational efficiency%
\footnote{This is further enhanced by moving to ``rational polynomial'' coordinates, in which $\vartheta$ is replaced by $y=\cos{\vartheta}$ (cf. Sec. 1.6 of Ref. \cite{Wiltshire+2009}).} %
as well as physical clarity. This series of coordinate transformations also makes it formally clear that the interior metric, due to analytic continuation, is simply given by the exterior one, with the timelike and spacelike coordinates legitimately switching character.


\subsection{Zero Angular Momentum Null Expansions}
\label{sec:SecIID_PNCs}

In this section, we are interested in obtaining the expansion scalars associated with two special congruences of zero angular momentum null geodesics. In Sec. \ref{sec:SecVD_PNEs_Results}, we will use these to explore the cosmology in the BH interior, using a representative set of BH models. Each scalar captures the fractional rate of change of cross-sectional area of the congruence (see, e.g., Sec. 2.4.8 of Ref. \cite{Poisson2004}), and corresponds to the trace of its extrinsic curvature.

For our purposes, it is useful to work with an orthonormal complex null tetrad \cite{Newman+1961, Kinnersley1969}. The Newman-Penrose (NP) tetrad, $\{\ell_+^{\dagger\mu}, \ell_-^{\dagger\mu}, \zeta^\mu, \bar{\zeta}^\mu\}$, has two real legs, $\ell_+^{\dagger\mu}$ and $\ell_-^{\dagger\mu}$, and two complex ones, with $\bar{\zeta}^\mu$ denoting the complex-conjugate of $\zeta^\mu$. The tetrad orthonormality relations can be read off from
\begin{equation} \label{eq:NP_Tetrad_Normalization}
\mathscr{g}^{\mu\nu} = -2\ell_+^{(\dagger\mu}\ell_-^{\dagger\nu)} + 2\zeta^{(\mu}\bar{\zeta}^{\nu)}\,.
\end{equation}

For region I, in BL coordinates, we can write the NP tetrad explicitly as, 
\begin{align} \label{eq:NP_Null_Tetrad_RegionI}
\ell_\pm^{\dagger\mu} =&\ \sqrt{\frac{\Pi}{2\Delta\Sigma}}\left[1, \pm\frac{1}{\sqrt{\Pi}}\frac{\Delta}{\sqrt{g}}, 0, \frac{2aF}{\Pi}\right]\,, \\
\zeta^\mu =&\ \frac{(\sqrt{A} - \mathrm{i}a\cos{\vartheta})}{\sqrt{2\Sigma}}\left[0, 0, \frac{1}{\sqrt{\Sigma}}, \frac{\mathrm{i}}{\sqrt{\Pi/\Sigma}\sin{\vartheta}}\right]\,. \nonumber 
\end{align}
The vector fields $\ell_\pm^{\dagger\mu}$ describe zero angular momentum (ZAM) null congruences, $(\ell_\pm^{\dagger})_{\vartheta} = (\ell_\pm^{\dagger})_{\varphi} = 0$, and are related to the generators of the PNCs introduced in eq. \ref{eq:PNC_Vector_Fields_BL}, $\ell^\mu_\pm$, through a series of null tetrad rotations (see Sec. 8 (g) of Ref. \cite{Chandrasekhar1985}). 

With the above, the expansions of the outgoing and ingoing ZAM null congruences, $\ell_\pm^\dagger$, are defined as,%
\footnote{Denoted in Ref. \cite{Hawking1968} by $\rho$ and $\mu$, upto signs.}
\begin{align}
\Theta_+ = (\ell_+^\dagger)_{\mu; \nu} \zeta^\mu \bar{\zeta}^\nu\,;\quad
\Theta_- = (\ell_-^\dagger)_{\mu; \nu} \bar{\zeta}^\mu \zeta^\nu\,,
\end{align}
where the semicolons denote covariant derivatives. In the relevant $\vartheta\varphi-$sector, we find $(\ell_+^\dagger)_{[\mu; \nu]} = (\ell_-^\dagger)_{[\mu; \nu]} = 0$, so that the above wonderfully becomes
\begin{align}
\Theta_\pm = (\ell_\pm^\dagger)_{(\mu; \nu)} \zeta^{(\mu}\bar{\zeta}^{\nu)}\,,
\end{align}
from which we can explicitly see that each ZAM null expansion is the covariant derivative of the congruence projected onto the $2-$sphere (see eq. \ref{eq:NP_Tetrad_Normalization}; See also, e.g., Sec. 2.4 of Ref. \cite{Poisson2004}). Furthermore, for each component we have $|(\ell_+^\dagger)_{\mu; \nu}| = |(\ell_-^\dagger)_{\mu; \nu}|$. Therefore, the null expansions are equal in magnitude $\Theta_+ = -\Theta_-$, and can be read off from (region I)
\begin{align} \label{eq:Null_Expansion_RegI}
\Theta_+\Theta_- = -\frac{\Delta}{32\Sigma g}\left(\frac{\partial_r \Pi}{\Pi}\right)^2 =  -\frac{\mathscr{g}^{rr}}{8}\left(\partial_r\ln{[\sqrt{\sigma}]}\right)^2\,,
\end{align}
where $\sigma$ is the determinant of the induced metric on a BL coordinate $2-$sphere,
\begin{equation}
\sigma = \mathscr{g}_{\vartheta\vartheta}\mathscr{g}_{\varphi\varphi} = \Pi\sin^2{\vartheta}\,.
\end{equation}
The null expansions are explicitly given in terms of the cross-sectional area of the congruence, $\delta S=\sqrt{\sigma}\mathrm{d}\vartheta\mathrm{d}\varphi$, as 
\begin{equation} \label{eq:PNEs}
|\Theta_+| = |\Theta_-| = \frac{\sqrt{\mathscr{g}^{rr}}\partial_r(\delta S)}{2\sqrt{2}(\delta S)}\,.
\end{equation}
For the Schwarzschild metric, we find $\Theta_\pm = \pm\sqrt{r-2M}/\sqrt{2r^3}$. Thus, for $r > 2M$, we can see that the outgoing radial null congruence is expanding and the ingoing radial null congruence is converging. For completeness, we note that, in spherically-symmetric spacetimes, $H = \sqrt{-8\Theta_+\Theta_-}$ is independent of $\vartheta$, and corresponds to the mean curvature of the coordinate $2-$sphere.

For region II, using the interior BL coordinates (Sec. \ref{sec:SecIIC_Interior_Coordinates}), we can write the NP tetrad explicitly as, 
\begin{align} \label{eq:NP_Null_Tetrad_RegionII}
\ell^{\dagger\breve{\mu}}_\pm =&\ \sqrt{\frac{\Pi}{-2\Delta\Sigma}}\left[\frac{1}{\sqrt{\Pi}}\frac{\Delta}{\sqrt{g}}, \pm 1, 0, \pm\frac{2aF}{\Pi}\right]\,, \\
\zeta^{\breve{\mu}} =&\ \frac{(\sqrt{A} - \mathrm{i}a\cos{\vartheta})}{\sqrt{2\Sigma}}\left[0, 0, \frac{1}{\sqrt{\Sigma}}, \frac{\mathrm{i}}{\sqrt{\Pi/\Sigma}\sin{\vartheta}}\right]\,.\nonumber 
\end{align}
The null expansions are equal, $\Theta_+ = \Theta_-$, and both negative. This indicates that the outgoing and the ingoing ZAM null congruences are both converging with time. Remarkably, 
\begin{align}
\Theta_+\Theta_- = -\frac{\Delta}{32\Sigma g}\left(\frac{\partial_r \Pi}{\Pi}\right)^2 =  -\frac{\mathscr{g}^{rr}}{8}\left(\partial_r\ln{[\sqrt{\sigma}]}\right)^2\,,
\end{align}
which is the same as eq. \ref{eq:Null_Expansion_RegI}. For the Schwarzschild metric, we find $\Theta_\pm = -\sqrt{2M-r}/\sqrt{2r^3}$. 


\section{Matter Comoving Frame}
\label{sec:SecIII_Comoving_Frame}

We have thus far discussed the properties of the axisymmetric metric \textit{ansatz} \eqref{eq:Stationary_Metric}. We now turn to a systematic consideration of the properties of the matter that generates the spacetime metric, beginning with the determination of the matter rest-frame.

We define the comoving or rest frame of the self-gravitating matter that generates the stationary spacetime as being given by an orthonormal tetrad, $\{e_{(a)}^\mu\ (a = 0\!-\!3)\}$, adapted to a \textit{timelike} congruence in which there are no energy or momentum fluxes. That is, we will orient the tetrad legs such that the matter energy-momentum-stress (EMS) tensor, $\mathscr{T}_{\mu\nu}$, appears diagonal in this frame. If such a frame exists, then it can be obtained by a local boost along with a local $\mathrm{SO}(3)-$rotation from the coordinate basis frame, $\{\delta^\mu_\alpha/\sqrt{|\mathscr{g}_{\alpha\alpha}|}\ (\alpha = 0\!-\!3)\}$.

In the comoving frame, we identify the rest-frame energy density, $\epsilon$, and principal pressures, $p_i\ (i = 1\!-\!3)$, of the background matter content as
\begin{align} \label{eq:Rest_Frame_Matter_Content_Def}
\epsilon =&\ \mathscr{T}_{(t)(t)} := \mathscr{T}_{\mu\nu}e^\mu_{(t)}e^\nu_{(t)}\,, \\
p_i =&\ \mathscr{T}_{(i)(i)} := \mathscr{T}_{\mu\nu}e^\mu_{(i)}e^\nu_{(i)}\,. \nonumber
\end{align}
In the above, $e^\mu_{(t)}$ denotes the timelike leg of the tetrad, corresponding to the normalized $4-$velocity of the matter, and $e^\mu_{(i)}$ denote the three normalized spacelike legs that are orthogonal to it.

Matter that is minimally-coupled to Einstein-Hilbert gravity satisfies the Einstein equations (see, e.g., Ref. \cite{Blau2023}), which relate the curvature of spacetime to the matter EMS tensor as $\mathscr{G}_{\mu\nu} = 8\pi\mathscr{T}_{\mu\nu}$. In such theories, the properties of the matter given in eq. \ref{eq:Rest_Frame_Matter_Content_Def} are equivalently written as
\begin{align} \label{eq:Rest_Frame_Matter_Content_Einstein}
\epsilon =&\ \mathscr{G}_{\mu\nu}e^\mu_{(t)}e^\nu_{(t)}/(8\pi)\,, \\
p_i =&\ \mathscr{G}_{\mu\nu}e^\mu_{(i)}e^\nu_{(i)}/(8\pi)\,. \nonumber
\end{align}
Here $\mathscr{G}_{\mu\nu}$ is the Einstein tensor, which is given in terms of the Ricci curvature tensor, $\mathscr{R}_{\mu\nu}$, and its trace, $\mathscr{R} = \mathscr{g}^{\mu\nu}\mathscr{R}_{\mu\nu}$, as $\mathscr{G}_{\mu\nu} = \mathscr{R}_{\mu\nu} - \mathscr{g}_{\mu\nu}\mathscr{R}/2$. 

Equations \ref{eq:Rest_Frame_Matter_Content_Def} and \ref{eq:Rest_Frame_Matter_Content_Einstein} reveal how the \textit{orthonormalized} eigenvectors of the Einstein tensor of the stationary spacetime define the matter comoving frame.%
\footnote{Note that the orthonormalization condition implies that these are also eigenvectors of the metric tensor.} %
Furthermore, the rest-frame energy density is the eigenvalue of the Einstein tensor corresponding to the timelike eigenvector. The remaining eigenvalues yield the rest-frame principal pressures.

When the matter content has several different components (e.g., a scalar field, an electromagnetic field, fluid matter), no restrictions are imposed on the form of the field-couplings between them, i.e., they can either be minimally- or nonminimally-coupled to each other. The EMS tensor in the Einstein equations then is the ``total'' EMS tensor, envisioned as having contributions from all matter components. We interpret the energy density and the principal pressures in eq. \eqref{eq:Rest_Frame_Matter_Content_Einstein} then as the total rest-frame values.

In stationary regions of the spacetime (where $\Delta > 0$; e.g., region I), the Einstein tensor for the stationary metric \eqref{eq:Stationary_Metric} has only one independent off-diagonal component, $\mathscr{G}_{t\varphi}$.%
\footnote{This is ensured by the choice of the conformal factor $X = \Sigma$. As an aside, we note that for the metric \eqref{eq:Stationary_Metric_X} with general $X\neq\Sigma$, one also obtains $\mathscr{G}_{r\vartheta} \neq 0$. In this case, two of the eigenvectors of the rest-frame do not, in general, lie precisely along $E_1$ and $E_2$ \eqref{eq:Einstein_Eigenvectors} but lie in their span instead.
\label{fn:X_neq_Sigma}} %
This enables finding, straightforwardly, its eigenvectors to be, in general,
\begin{align} \label{eq:Einstein_Eigenvectors}
\begin{alignedat}{3}
& E^\mu_0 \propto\ \left[1, 0 ,0, \Omega\right]\,;\
&& E^\mu_3 \propto\ \left[1, 0 ,0, \chi\right]\,, \\
& E^\mu_1 \propto\ \left[0, 1, 0, 0\right]\,;\
&& E^\mu_2 \propto\ \left[0, 0, 1, 0\right]\,.
\end{alignedat} 
\end{align}
In particular, we will require, reasonably, that the timelike eigenvector of the above corresponds to the matter $4-$velocity.

Since in stationary regions $E_1$ and $E_2$ are spacelike, we see that the matter configuration obeys the underlying symmetries of the spacetime geometry, i.e., its $4-$velocity is always described by (a quasi-Killing congruence \cite{Iyer+1993})
\begin{equation} \label{eq:Stationary_Observer}
e_{(t)}^\mu \propto E_0^\mu \propto \left[1, 0 ,0, \Omega(r, \vartheta)\right]\,.
\end{equation}
As indicated above, in general, $\Omega$ need not be constant on spherical (constant$-r$) surfaces, i.e., the matter can be differentially-rotating. However, as we will see, $\Omega$ is independent of $\vartheta$ in a large class of special spacetimes.

The complete rest frame is then defined concretely as being 
\begin{align} \label{eq:Comoving_Frame_Stationary_regionI}
\begin{alignedat}{3}
& e^\mu_{(t)} =\ \frac{1}{\sqrt{-N(\Omega)}}\left[1, 0 ,0, \Omega\right]\,;\
&& e^\mu_{(\varphi)} =\ \frac{1}{\sqrt{+N(\chi)}}\left[1, 0 ,0, \chi\right]\,;\\
& e^\mu_{(r)} =\ \frac{1}{\sqrt{+\mathscr{g}_{rr}}}\left[0, 1, 0, 0\right]\,;\
&& e^\mu_{(\vartheta)} =\ \frac{1}{\sqrt{+\mathscr{g}_{\vartheta\vartheta}}}\left[0, 0, 1, 0\right]\,.
\end{alignedat} 
\end{align}
In the above, we have made explicit the requirement that $e_{(t)}$ and $e_{(\varphi)}$ be timelike ($N(\Omega) < 0$) and spacelike ($N(\chi) > 0$) respectively. This simply implies that $\Omega_- < \Omega < \Omega_+$ as well as that either $\chi < \Omega_-$ or $\chi > \Omega_+$. For the special case of $\Omega = \Omega_{\mathrm{Z}}$, we can simply set $e^\mu_{(\varphi)} = \delta^\mu_\varphi/\sqrt{\mathscr{g}_{\varphi\varphi}}$ (cf. Ref. \cite{Bardeen+1972}).

Furthermore, due to orthonormality ($\mathscr{g}_{\mu\nu}e^\mu_{(t)}e^\nu_{(\varphi)} = 0$), $\chi$ is fixed by $\Omega$ as
\begin{align} \label{eq:Chi}
\chi(\Omega) =&\ - \frac{\mathscr{g}_{tt} + \Omega\mathscr{g}_{t\varphi}}{\mathscr{g}_{t\varphi} + \Omega\mathscr{g}_{\varphi\varphi}} = \Omega_- - \frac{(\Omega_+ - \Omega_-)(\Omega - \Omega_-)}{2(\Omega_{\mathrm{Z}}-\Omega)} \nonumber \\
=&\ \Omega_+ + \frac{(\Omega_+ - \Omega_-)(\Omega_+ - \Omega)}{2(\Omega-\Omega_{\mathrm{Z}})}\,. 
\end{align}
Clearly, if $\Omega_- < \Omega < \Omega_{\mathrm{Z}}$, we automatically have $\chi < \Omega_-$. Similarly, if $\Omega_{\mathrm{Z}} < \Omega < \Omega_+$, we automatically have $\chi > \Omega_+$. 

Thus, finding the matter comoving frame in region I reduces to finding a real root $\Omega_- < \Omega < \Omega_+$ to the quadratic equation $\mathscr{G}_{(t)(\varphi)} = 0$,
\begin{align} \label{eq:Omega_Quadratic}
& \left[\mathscr{g}_{t\varphi}\mathscr{G}_{\varphi\varphi} - \mathscr{g}_{\varphi\varphi}\mathscr{G}_{t\varphi}\right]\Omega^2 
+ \left[\mathscr{g}_{tt}\mathscr{G}_{\varphi\varphi} - \mathscr{g}_{\varphi\varphi}\mathscr{G}_{tt}\right]\Omega 
\nonumber \\
&\ + \left[\mathscr{g}_{tt}\mathscr{G}_{t\varphi} - \mathscr{g}_{t\varphi}\mathscr{G}_{tt}\right] = 0\,.
\end{align}
In case such a root does not exist, we can conclude that matter described by the metric above \eqref{eq:Stationary_Metric} flows on achronal orbits (those that do not contain any points connected by a timelike path; cf. Ref. \cite{Graham+2007}). 

In nonstationary regions of the maximally-extended spacetime (e.g., region II), we can perform an identical analysis, using the interior BL-like coordinates introduced above (Sec. \ref{sec:SecIIC_Interior_Coordinates}). In this way, we find that the matter orthonormal rest-frame is given as,
\begin{align} \label{eq:Comoving_Frame_Stationary_regionII}
\begin{alignedat}{3}
& e^{\breve{\mu}}_{(\tau)} =\ \frac{1}{\sqrt{-\mathscr{g}_{rr}}}\left[-1, 0, 0, 0\right]\,;\
&& e^{\breve{\mu}}_{(\phi)} =\ \frac{1}{\sqrt{+N(\chi)}}\left[0, 1, 0, \chi\right]\,;\\
& e^{\breve{\mu}}_{(z)} =\ \frac{1}{\sqrt{+N(\Omega)}}\left[0, 1, 0, \Omega\right]\,;\
&& e^{\breve{\mu}}_{(\vartheta)} =\ \frac{1}{\sqrt{+\mathscr{g}_{\vartheta\vartheta}}}\left[0, 0, 1, 0\right]\,,
\end{alignedat} 
\end{align}
where $e_{(\tau)}$ is the matter timelike $4-$velocity and $e_{(\vartheta)}$ is a spacelike leg. Moreover, $e_{(z)}$ and $e_{(\phi)}$ are assuredly spacelike in this region since they are both proportional to the general Killing vector $\mathbf{K}$ (see discussion below eq. \ref{eq:Stationary_Metric_iBL}). In the above, we have used the interior metric in the form \eqref{eq:Interior_Metric_RI}, so that $\mathscr{g}_{rr}$ and $\mathscr{g}_{\vartheta\vartheta}$ continue to denote the BL metric in \textit{region I}, i.e., $\mathscr{g}_{rr}=(\Sigma/\Delta)g$ and $\mathscr{g}_{\vartheta\vartheta}=\Sigma$. Similarly, $N(\Omega)$ is also given in terms of the BL metric functions in region I via eq. \ref{eq:Killing_Norm}.

This shows us that in the nonstationary regions of spacetime (e.g., between two horizons), the matter is comoving with the interior cosmology. The rest-frame angular frequency $\Omega$ now determines the precession frequency of the spatial triad of the matter rest-frame. Thus, it plays a fundamentally different role from that in the stationary regions of the BH spacetime, where it captures the matter orbital angular frequency about the spin axis.

Putting everything together, the rest-frame energy density and principal pressures are given everywhere in the maximally-extended spacetime as
\begin{align} \label{eq:Rest_Frame_Matter_Properties}
\epsilon =&\ 
\begin{cases}
+\mathscr{G}_{(t)(t)}/(8\pi)\,, & \Delta > 0 \\
-\mathscr{G}_{(r)(r)}/(8\pi)\,, & \Delta < 0
\end{cases}\,, \\
p_n =&\ 
\begin{cases}
+\mathscr{G}_{(r)(r)}/(8\pi)\,, & \Delta > 0 \\
-\mathscr{G}_{(t)(t)}/(8\pi)\,, & \Delta < 0
\end{cases}\,, \nonumber \\
p_\vartheta =&\ \mathscr{G}_{(\vartheta)(\vartheta)}/(8\pi)\,, \nonumber \\
p_\varphi =&\ \mathscr{G}_{(\varphi)(\varphi)}/(8\pi)\,. \nonumber
\end{align}
In the above, on the right-hand side, $\mathscr{G}_{(a)(b)}$ always denotes the rest-frame projected Einstein tensor in \textit{region I}. This is because we have used the interior metric in the form \eqref{eq:Interior_Metric_RI}. We note that $p_n$ denotes the principal pressure in directions normal to $2-$spheres everywhere in the spacetime. Therefore, we will henceforth refer to it as the normal pressure. In spherically-symmetric spacetimes, by symmetry, the two tangential pressures, $p_\vartheta$ and $p_\varphi$, are always equal.

In regions of spacetime where the rest-frame energy-density is non-negative $\epsilon \geq 0$, the statements of the various classical energy conditions can be reduced into simpler conditions on the normal ($\omega_n$), polar ($\omega_\vartheta$), and azimuthal ($\omega_\varphi$) equations-of-state (EoSs),
\begin{equation} \label{eq:EoS}
\omega_n = p_n/\epsilon\,;\ 
\omega_\vartheta = p_\vartheta/\epsilon\,;\ 
\omega_\varphi = p_\varphi/\epsilon\,.
\end{equation}
For example, the NEC and the WEC are satisfied if $\omega_i \geq -1$. The DEC is satisfied if $|\omega_i| \leq 1$. The SEC is satisfied if $\omega_i \geq -1$ as well as $\Sigma_i \omega_i \geq -1$. We direct the reader to \ref{app:AppD_Energy_Conditions} for a brief summary of the various energy conditions discussed here.


\subsection{Matter on Horizons}
\label{sec:SecIIIA_Horizons}

We now consider the properties of matter at the future horizon. First, we restrict ourselves to spherically-symmetric spacetimes \eqref{eq:Static_Metric}. The components of the Einstein tensor for these spacetimes are given explicitly in \ref{app:AppE_Sph_Symm_EEs}. 

The following combination of the energy density and the normal pressure is particularly instructive,
\begin{align} \label{eq:Radial_Pressure_v1}
\epsilon + p_n = \pm\frac{\hat{\mathscr{G}}_{(t)(t)} + \hat{\mathscr{G}}_{(r)(r)}}{8\pi} = \pm\frac{1}{4\pi}\left(1 - \frac{2m}{R}\right)\frac{\partial_r \psi}{R\partial_r R}\,,
\end{align}
where the upper (lower) sign applies in regions where $\mathbf{T} = \partial_t$ is timelike (spacelike).

Since the horizons for the spherically-symmetric spacetimes always occur at $R(r) = 2m(r)$, eq. \ref{eq:Radial_Pressure_v1} immediately reveals that matter in the limit of approach to the horizon always satisfies (see also Refs. \cite{Visser1992, Medved+2004, Lobo+2020})
\begin{equation} \label{eq:Horizon_NEC}
\epsilon + p_n = 0, \quad \omega_n=-1.
\end{equation}
The equation above is quite remarkable. It is a statement about the properties of matter, at a horizon, in its own \textit{rest frame}, and can, therefore, be used to set up several ``no-go'' theorems. Matter content whose normal EoS \eqref{eq:EoS} does not satisfy $\omega_n = -1$ cannot form static BH spacetimes. 

As we show in \ref{app:AppF1_Scalar_Fields}, a massless minimally-coupled scalar field in a spherically-symmetric spacetime has $\omega_n = +1$. Thus, it cannot describe a static BH solution (cf. also Ref. \cite{Bekenstein1972} for a related no-go theorem). Indeed, the static and spherically symmetric solution to the Einstein-Klein-Gordon system is given by the Wyman-Janis-Newman-Winicour metric \cite{Wyman1981, Janis+1968, Virbhadra1997}, which describes a naked singularity and not a black hole. On the other hand, we discuss in \ref{app:AppF2_EM_Fields} how $\omega_n = -1$ for arbitrary spherically symmetric configurations of electromagnetic fields. Indeed, the unique electrovacuum solution to the Einstein-Maxwell system, namely the Reissner-Nordstr{\"o}m metric (cf., e.g., Ref. \cite{Wald1984}), can describe BH solutions. Finally, we note that the Joshi-Malafarina-Narayan-1 naked singularity metric \cite{Joshi+2011} describes a spherically symmetric configuration of anisotropic fluid matter with $\omega_n = 0$. Equation \ref{eq:Horizon_NEC} then indicates that this matter model cannot be used to form static BH solutions. 

Importantly, as can be seen from eq. \ref{eq:Rest_Frame_Matter_Properties}, we note that eq. \ref{eq:Horizon_NEC} also holds for matter on the horizons of the broad class of stationary and axisymmetric metric that we consider here \eqref{eq:Stationary_Metric}. See also the related discussion in Sec. 5.4.2 of Ref. \cite{Poisson2004}, Sec. 6.3.1 of Ref. \cite{Hawking+2010}, and in Ref. \cite{Jacobson1995}. Thus, similar no-go theorems can be set up even in stationary and axisymmetric spacetimes. It is interesting to consider the implications of such no-go theorems for the weak cosmic censorship conjecture \cite{Penrose2002, Hawking+1973, Joshi2007}. 

These no-go theorems only impose restrictions on the types of matter that are permitted to exist \textit{on} a BH horizon. In particular, they do not restrict self-gravitating matter in non-empty BH spacetimes from having $\omega_n \neq -1$ either in the BH exterior or in its interior.

We emphasize that these no-go theorems apply only to stationary (equilibrium) solutions. This becomes immediately clear upon considering the spacetime generated by the continual collapse (Oppenheimer-Snyder or Lema{\^i}tre-Tolman-Bondi \cite{Oppenheimer+1939, Lemaitre1997, Tolman1934, Bondi1947, Singh+1994}) or continual expansion (Friedman-Lema{\^i}tre-Robertson-Walker \cite{Friedman1999, Lemaitre1931, Lemaitre1997, Robertson1935, Walker1937}) of a spherically-symmetric configuration of self-gravitating dust ($\omega_i = 0$).


\subsection{Degenerate Spacetimes}
\label{sec:SecIIIB_Degenerate_Spacetimes}

The line element for a very special class of spherically-symmetric spacetimes, whose metric functions take values $\psi(r) = 0$ and $R(r) = r$,%
\footnote{For spacetimes with $\psi = 0$, we can trivially set $R = r$.} 
is given as,
\begin{align}  \label{eq:Degenerate_Spacetimes}
\mathrm{d}s^2 =-\left(1-\frac{2m}{r}\right)\mathrm{d}t^2 + \left(1-\frac{2m}{r}\right)^{-1}~\mathrm{d}r^2 + r^2~\mathrm{d}\Omega_2^2\,.
\end{align}
Since these spacetimes are described by a single metric function $m(r)$, we will refer to them as degenerate spacetimes. These spacetimes satisfy $\hat{\mathscr{g}}_{tt}\hat{\mathscr{g}}_{rr} = - 1$ in \textit{areal-radial} coordinates, and have been analyzed in Ref. \cite{Jacobson2007}. 

Well-known examples of degenerate spacetimes include the Schwarzschild and Reissner-Nordstr{\"o}m BHs of GR as well as several ``regular'' BH models that have been proposed to address the issue of the curvature singularity (cf. Refs. \cite{Bardeen1968, Ayon-Beato+1998, Dymnikova2004, Hayward2006, Zhou+2023}).

Since $\psi(r) = 0$, we can see from eq. \ref{eq:Radial_Pressure_v1} that degenerate spacetimes contain matter that satisfies $p_n = -\epsilon$ everywhere, not just at the event horizon. More completely, in region I, we can obtain 
\begin{equation}
\hat{\mathscr{T}}_{(t)(t)} = -\hat{\mathscr{T}}_{(r)(r)} = \frac{\partial_r m}{4\pi r^2}\,;\ 
\hat{\mathscr{T}}_{(\vartheta)(\vartheta)} = \hat{\mathscr{T}}_{(\varphi)(\varphi)} = -\frac{\partial_r^2 m}{8\pi r}\,.
\end{equation}
All of the energy conditions reduce to conditions on the metric function $m$ and its first two derivatives. For spacetimes with $\partial_r m > 0$, the validity of the energy conditions can be determined by simply examining the behaviour of $\omega_\vartheta = -r\partial_r^2m/(2\partial_r m)$. All of the above applies equally well in region II with the formal replacement $r \leftrightarrow \tau$.

As an illustrative example, let us consider the class of regular Hayward-like BH spacetimes (cf. Refs. \cite{Hayward2006, Zhou+2023}), whose mass functions are given as
\begin{equation}
m(r) = \frac{M r^p}{r^p + \alpha L^{p-q}}\,,
\end{equation}
where $L, p, q$, and $\alpha$ are some positive constants. It can be checked that the rest-frame energy density for these spacetimes is assuredly positive-definite, $\epsilon > 0$. The tangential equation of state for these BH models is given as,
\begin{equation} \label{eq:Omega_th_Gen_Hayward}
\omega_\vartheta(r) = \frac{1 + p}{2} - \frac{p \alpha L^{p-q}}{r^p + \alpha L^{p-q}}\,,
\end{equation}
so that, at the center ($r=0$), we can write
\begin{equation} \label{eq:Center_EoS_Regular}
\omega_\vartheta(0) = (1-p)/2\,.
\end{equation}
Thus, degenerate Hayward-like BH spacetimes with $p>3$ generically violate all of the classical energy conditions. For $p\leq 3$, energy condition violating regions may still persist \eqref{eq:Omega_th_Gen_Hayward}. For the original Hayward BH metric \cite{Hayward2006} in particular, for which $(p, q, \alpha) = (3, 1, 2M)$, as we shall see in Sec. \ref{sec:SecIV_Nonspinning_BH_Properties}, all energy conditions barring the DEC are satisfied everywhere.

Furthermore, the rest-frame angular velocity of the matter in the \textit{spinning} counterparts \eqref{eq:Stationary_Metric} of degenerate spacetimes \eqref{eq:Degenerate_Spacetimes}, can be obtained, in the stationary regions of spacetime, by solving eq. \ref{eq:Omega_Quadratic}. It is found to be (cf. Refs. \cite{Azreg-Ainou2014a, Azreg-Ainou2014b, Azreg-Ainou2014c})
\begin{equation} \label{eq:Omega_PNC}
\Omega(r) = \frac{a}{A(r)+a^2} = \Omega_{\mathrm{PNC}}(r)\,,
\end{equation}
which is independent of $\vartheta$. In the above, we have introduced $\Omega_{\mathrm{PNC}}$ to denote the angular frequency of the principal null congruences in stationary regions of spacetime (see equation \ref{eq:PNC_Vector_Fields_BL}),
\begin{equation}
\Omega_{\mathrm{PNC}} = \ell_\pm^\varphi/\ell_\pm^t .
\end{equation}
We can also check that the matter $4-$velocity \eqref{eq:Comoving_Frame_Stationary_regionI} is assuredly timelike in region I since $N(\Omega_{\mathrm{PNC}}) = -\Delta\Sigma/(A+a^2)^2$. 

Note also that, unsurprisingly, all spacetimes that satisfy the special requirement (see \ref{app:AppG_AA_Comparison}) 
\begin{equation} \label{eq:OmegaPNC_Condition}
R^2(r) = c_2 r^2 + c_1 r + c_0\,;\ \  g(r) = 1\,,
\end{equation}
also contain rigidly rotating matter with $\Omega = \Omega_{\mathrm{PNC}}$. In such spacetimes where the matter angular velocity matches that of the PNCs, the matter rest-frame corresponds to the Carter tetrad of the spacetime \cite{Carter1968, Znajek1977}.


\section{Properties of Matter in Nonspinning Black Hole Spacetimes}
\label{sec:SecIV_Nonspinning_BH_Properties}


\begin{table*}[!ht]
\begin{center}
\caption{
\textit{Metric functions for popular spherically-symmetric black hole spacetimes.}
The Schwarzschild and the Reissner-Nordstr{\"o}m (RN) spacetimes describe the vacuum and the electrovacuum black hole (BH) solutions of general relativity (GR). The Hayward \cite{Hayward2006} and the $n=4$ modified Hayward metric by Zhou \& Modesto (ZM4; \cite{Zhou+2023}) describe regular BH spacetimes containing an anisotropic fluid. These are all degenerate BH spacetimes (Sec. \ref{sec:SecIIIB_Degenerate_Spacetimes}). The Gibbons-Maeda-Garfinkle-Horowitz-Strominger (GMGHS; \cite{Gibbons+1988, Garfinkle+1991}) BH metric is a solution to the low energy theory of the heterotic string. This theory is GR with additional matter fields, namely a scalar (``dilaton'') and an electromagnetic (EM) field. Finally, the Frolov BH metric \cite{Frolov2016} describes a regular BH spacetime, which modifies the redshift function $\psi(r)$ of the Hayward metric. The quantities $M$ and $Q$ denote the total mass and electromagnetic charge of the spacetime, and $L$ denotes a (de Sitter) length scale. For the GMGHS BH, its dilaton charge $D$ is given as $|D|=Q^2/(2M)$. The metric function $m(r)$ is the Hawking mass of a $2-$sphere of coordinate radius $r$, and $R(r)$ is the areal-radius of the sphere.}
\label{table:Known_Static_Solutions}
\renewcommand{\arraystretch}{1.5}
\centering
\begin{tabular}[t]{||l||l||l|l|l||l||}
\hline
Spacetime &  Matter Content  & $R(r)$ & $\mathrm{e}^{\psi(r)}$ & $m(r)$ & $m(R)$ \\
\hline
Schwarzschild & None & $r$ & $1$ & $M$ & $M$ \\ 
RN & EM Field & $r$ & $1$ & $M-\frac{Q^2}{2r}$ & $M-\frac{Q^2}{2R}$ \\
Hayward & Fluid & $r$ & $1$ & $\frac{Mr^3}{r^3+2ML^2}$ & $\frac{MR^3}{R^3+2ML^2}$ \\ 
ZM4 & Fluid & $r$ & $1$ & $\frac{Mr^4}{r^4+2L^4}$ & $\frac{MR^4}{R^4+2L^4}$ \\ 
GMGHS & Scalar, EM Field &
$\sqrt{r(r-Q^2/M)}$ & $(\partial_r R)^{-1}$ & $\frac{R}{2}\left[1-\left(1-\frac{2M}{r}\right)(\partial_r R)^2\right]$ & $M\left(\frac{\sqrt{D^2 + R^2} - D}{R}\right)\left(1 + \frac{D^2}{R^2}\right) - \frac{D^2}{2R}$\\
Frolov & Fluid & $r$ & $\frac{r^6+r_-^6}{r^6+r_+^4r_-^2}$ & $\frac{Mr^3}{r^3+2ML^2}$ & $\frac{MR^3}{R^3+2ML^2}$ \\  
\hline
\hline
\end{tabular}
\end{center}
\end{table*}

\begin{figure*}[ht!]
\centering
\includegraphics[width=2\columnwidth]{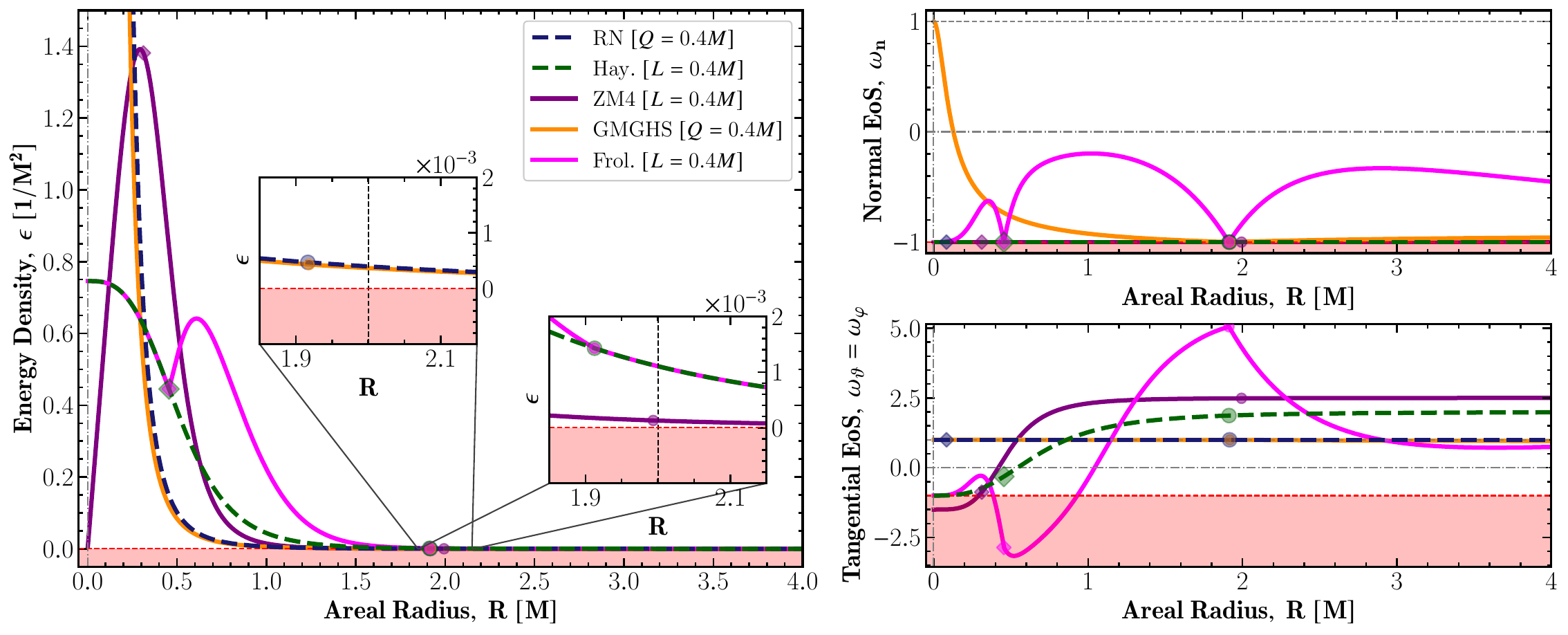}
\caption{
The left panel shows the distribution of the rest-frame energy-density $\epsilon$ as a function of the curvature or areal-radius $R$ in various non-spinning black hole (BH) spacetimes. The energy density is positive-definite everywhere in all of these spacetimes. Outer and inner horizons are located by circles and diamonds respectively. The upper inset zooms into the region close to the horizon of the electromagnetically-charged RN and GMGHS BHs. For the GMGHS metric, $\epsilon$ is continuous but non-differentiable across the horizon. The lower inset zooms into the near-horizon region of the regular Hayward, ZM4 and Frolov BH spacetimes. For the Frolov metric, $\epsilon$ is continuous but non-differentiable across the inner horizon. The top-right panel shows the normal equation of state (EoS) of the matter $\omega_n = p_n/\epsilon$ and the bottom-right panel shows its tangential EoS $\omega_\vartheta = p_\vartheta/\epsilon$ ($=\omega_\varphi$). Since $\epsilon > 0$ always, the weak energy condition (WEC) reduces to the requirement that $\omega_i \geq -1$. As can be seen from the red-shaded regions here, the WEC is violated only by the matter generating the regular ZM4 and Frolov spacetimes. All quantities are evaluated in the appropriate timelike comoving-frame of the matter (Sec. \ref{sec:SecIII_Comoving_Frame}).
}
\label{fig:Fig2_Energy_Density_Pressures_ECs}
\end{figure*}


We are finally in a position to use the concepts and methods described in the previous sections to study the properties of matter in black hole (BH) spacetimes. In this section, we consider spherically-symmetric nonspinning BHs, and in Sec. \ref{sec:SecV_Spinning_BH_Properties} we analyze the more general case of spinning BHs.

To guide the discussion, we focus on the six representative spherically-symmetric black hole (BH) spacetimes listed in Table \ref{table:Known_Static_Solutions}. This set of BH metrics allows us to cover several different aspects of the problem.

The Schwarzschild and Reissner-Nordstr{\"o}m (RN) spacetimes are solutions to the vacuum and electrovacuum equations of GR respectively (see, e.g., Ref. \cite{Misner+1973}). The Hayward metric \cite{Hayward2006} describes a regular BH spacetime, and can be interpreted as containing an anisotropic fluid (see, e.g., Ref. \cite{Doneva+2012}) with no heat fluxes or viscosity, i.e., its EMS tensor is given as
\begin{equation}
\mathscr{T}^{\mu\nu} = \epsilon e_{(t)}^\mu e_{(t)}^\nu + p_n e_{(r)}^\mu e_{(r)}^\nu + p_\vartheta e_{(\vartheta)}^\mu e_{(\vartheta)}^\nu + p_\varphi e_{(\varphi)}^\mu e_{(\varphi)}^\nu\,. \nonumber
\end{equation}
 Similar anisotropic fluids (with different equations of state, however) populate the ZM4 and the Frolov BH spacetimes that are discussed below as well as the spinning counterparts of these three spacetimes discussed in Sec. \ref{sec:SecV_Spinning_BH_Properties}. In the spherically-symmetric spacetimes discussed in this section, the two tangential pressures are equal $p_\varphi = p_\vartheta$.
 
The Schwarzschild, RN, and Hayward metrics are examples of degenerate spacetimes (Sec. \ref{sec:SecIIIB_Degenerate_Spacetimes}), and are described by their Hawking mass distribution, $m(r)$. Modified-Hayward metrics, also degenerate spacetimes with different mass functions, were studied in Ref. \cite{Zhou+2023}. We will refer to the $n=4$ modified-Hayward metric of Ref. \cite{Zhou+2023} as the Zhou-Modesto-4 (ZM4) metric. This metric contains no curvature singularity and remains geodesically-complete even when extended to negative $r$. Note the minor change in the denominator of the mass function (Table \ref{table:Known_Static_Solutions}) from eq. 13 there.

The Gibbons-Maeda-Garfinkle-Horowitz-Strominger (GMGHS; \cite{Gibbons+1988, Garfinkle+1991}) solution describes a BH spacetime containing a scalar (dilaton) and an electromagnetic field. These fields arise naturally in the low-energy limit of the heterotic string, with the gravitational Lagrangian still given by the Einstein-Hilbert term. The GMGHS metric is not degenerate due to a nontrivial $R(r)$, as shown in Table \ref{table:Known_Static_Solutions}. However, as with the previous metrics, it is described in coordinates in which $g(r) = (\mathrm{e}^\psi~\partial_r R)^2 = 1$. For further discussion on these spacetimes, see also Refs. \cite{Kocherlakota+2020, Kocherlakota+2021, EHTC+2022f, Vagnozzi+2023, Kocherlakota+2023}. 

Finally, yet another nontrivial modification to the Hayward metric, the regular Frolov BH metric \cite{Frolov2016}, modifies the redshift function, $\psi(r)$, as shown in Table \ref{table:Known_Static_Solutions}. We note that this metric is expressed in ``$g(r)\neq 1$ coordinates'' as (see eqs. 2.47, 2.50 of Ref. \cite{Frolov2016}),
\begin{align} \label{eq:Frolov_Metric}
\mathrm{d}s^2 =&\ -\left[\frac{r^6 + r_-^6}{r^6 + r_+^4 r_-^2}\right]^2\left(1-\frac{2Mr^2}{r^3 + 2ML^2}\right)\mathrm{d}t^2 \\ 
&\ + \left(1-\frac{2Mr^2}{r^3 + 2ML^2}\right)^{-1}\mathrm{d}r^2 + r^2\mathrm{d}\Omega_2^2\,, \nonumber
\end{align}
where $r_\pm(L)$ are the locations of the two horizons, i.e., the real positive roots of the cubic $r^3 - 2Mr^2 + 2L^2$. Putting the above \eqref{eq:Frolov_Metric} into coordinates in which $g(r) = 1$ requires numerically solving a differential equation (see discussion below eq. \ref{eq:Spherically_Symmetric_Metric_v1}). This is not convenient, so we leave the metric in the form given in Table \ref{table:Known_Static_Solutions}.

Fig. \ref{fig:Fig2_Energy_Density_Pressures_ECs} shows the energy-density $\epsilon$ (left panel), the normal equation of state (EoS; cf. eq. \ref{eq:EoS}) $\omega_n = p_n/\epsilon$ (top-right panel), and the tangential EoS $\omega_\vartheta = p_\vartheta/\epsilon$ ($=\omega_\varphi$) for all the spacetimes considered above, in the appropriate timelike comoving-frame of the matter. The matter content in all cases has positive-definite energy density. Furthermore, for all the degenerate spacetimes (RN, Hayward, ZM4), we see, as predicted (Sec. \ref{sec:SecIIIB_Degenerate_Spacetimes}), that $\omega_r = -1$ identically. In addition, all the models (except Schwarzschild) contain matter with anisotropic pressure, $\omega_n \neq \omega_\vartheta$.

Furthermore, for the singular BH models (RN, GMGHS), it can be seen then that their matter satisfies the dominant energy condition (DEC) since $|\omega_i| \leq 1$ (see \ref{app:AppD_Energy_Conditions} for a brief discussion on the energy conditions). This automatically implies the validity of the weak energy condition (WEC) as well as of the null energy condition (NEC). Additionally, since $\omega_r \geq -1$ and $\omega_\vartheta = \omega_\varphi > 0$, the strong energy condition (SEC) is also satisfied (since $\omega_r + 2\omega_\vartheta \geq -1$). Thus, the matter content in these spacetimes satisfies all classical energy conditions. The variations in the GMGHS EoSs from the RN ones are not clearly visible due to the scale. Note, in particular, that our conclusion regarding the GMGHS spacetime differs from that of Ref. \cite{Kar1997}, where it was found that the energy conditions are violated. The reason is that our definition of the rest-frame energy density corresponds correctly to the eigenvalue of the Einstein tensor corresponding to its timelike eigenvector \eqref{eq:Rest_Frame_Matter_Content_Einstein}.

The matter content in the regular Hayward BH spacetime satisfies the WEC, and therefore also the NEC, everywhere but not the DEC (since $\omega_\vartheta > 1$). This matter also satisfies the SEC outside the horizon (at the center, see that $\omega_r + 2\omega_\vartheta = -3$).

On the other hand, the matter content in the regular ZM4 BH spacetime violates all of the energy conditions, albeit only inside its inner horizon ($\omega_\vartheta < -1$). Outside the event horizon, it only violates the DEC. 

Finally, the matter content in the regular (\textit{non-degenerate}) Frolov BH spacetime again violates all of the energy conditions. Here, the energy-condition-violating region is not restricted to the interior of the inner horizon though it is still contained within the event horizon. In the exterior horizon geometry, however, only the DEC is violated.


\section{Properties of Matter in Spinning Black Hole Spacetimes}
\label{sec:SecV_Spinning_BH_Properties}

Thus far we have considered in detail the properties of the equilibrium matter that generates various popular nonspinning spacetimes. In this section, we present an analysis of the matter content that generates their spinning counterparts. In all cases here, the spinning spacetimes are described by the metric \textit{ansatz} \eqref{eq:Stationary_Metric}. For a specific axisymmetric spacetime, one requires as inputs the spherically-symmetric ``seed'' metric functions given in Table \ref{table:Known_Static_Solutions} above.

The spinning spacetimes we will consider here are as follows. The Kerr metric is the stationary generalization of the Schwarzschild metric and is a vacuum solution of the Einstein equations, used to describe the vacuum, spinning BHs of general relativity (GR). The Kerr-Newman (KN) metric \cite{Newman+1965b} is the spinning counterpart of the RN metric and is a solution to the Einstein-Maxwell equations, used to describe electromagnetically charged, spinning BHs in GR. The KN metric was originally obtained via the Newman-Janis (NJ) algorithm \cite{Newman+1965}.

The Kerr-Hayward (KH) metric was obtained from the Hayward metric in Ref. \cite{Bambi+2013}, via the original NJ algorithm. An analysis of the energy conditions for this spacetime was also presented there, with which our results are consistent. This metric can be used to describe spinning regular BHs containing an anisotropic fluid. The Azreg-A{\"i}nou (AA) algorithm produces the same metric \cite{Azreg-Ainou2014c}. 

We introduce the Kerr-Zhou-Modesto-4 (KZM4) metric here, which can also similarly be obtained via either the NJ or AA algorithm, using the ZM4 metric as the seed metric. We remind the reader that all of the aforementioned spacetimes are special, and belong to the class of degenerate spacetimes discussed in Sec. \ref{sec:SecIIIB_Degenerate_Spacetimes}.

The Kerr-Sen (KS) spacetime \cite{Sen1992} describes the stationary generalization of the GMGHS metric, and contains a scalar (dilaton), an electromagnetic, and a pseudoscalar (axion) field. It was shown in Ref. \cite{Yazadjiev2000} that this metric can be obtained as a product of the original NJ algorithm. It can also be obtained via the AA algorithm. It is interesting to note that while the existence of a Carter constant for null geodesics for this metric \cite{Hioki+2008} leads to the expectation that it is of Petrov-Pirani type D, it has been shown in Ref. \cite{Burinskii1995} that it is actually of type I. 

Finally, we introduce the Kerr-Frolov (KF) spacetime here as being given by the metric \eqref{eq:Stationary_Metric}, with the seed metric functions given in Table \ref{table:Known_Static_Solutions}. Note that neither the KS nor the KF spacetimes are degenerate spacetimes. Furthermore, the nonspinning seed metrics for all but the KF spacetime are written in ``$g=1$'' coordinates. The Frolov metric does not admit a simple analytic form in coordinates in which $g(r)=1$. Strikingly, the KF BH metric, obtained in this way, contains a curvature singularity. We remind the reader that the seed metric was perfectly regular.

The simplicity in the construction of the KF metric in particular highlights the potential value of the AA algorithm. We will first discuss the properties of matter in the set of spacetimes for which the rest-frame angular velocity of the matter is given as $\Omega =\Omega_{\mathrm{PNC}}$ and return later to the KF metric, for which $\Omega \neq \Omega_{\mathrm{PNC}}$.


\subsection{Causality Violating Regions}
\label{sec:SecVA_CTCs}

In the discussion below, we will exclusively employ the areal-radii $\mathscr{R}_{\mathscr{A}}(r)$ of $2-$spheres of coordinate radius $r$, to enable meaningful comparisons of spatially-varying quantities across different spacetimes. This is because the areal-radius is a gauge-invariant (coordinate-invariant) quantity \eqref{eq:Areal_Radius}.

Furthermore, the areal-radius is defined only in regions that are devoid of closed timelike curves (CTCs), where $\Pi > 0$ (see Footnote \ref{fn:CTCs}), since $\mathscr{R}_{\mathscr{A}}$ involves an integral over $\sqrt{\Pi}$. We identify the causality-violating region to lie within the largest surface $r=r_{\mathrm{CTC}}$ on which $\Pi(r_{\mathrm{CTC}}, \vartheta) = 0$ for some $\vartheta$. It is possible that $\Pi$ becomes positive again inside the outermost $\Pi = 0$ surface. We will nevertheless disregard these regions, for simplicity. 

As a reminder, we note that when the spacetime contains a curvature singularity, it is located in the equatorial plane at $r=r_0$ such that $\Sigma(r_0, \pi/2) = 0$ or equivalently, $R(r_0) = 0$.

For the Kerr metric, we have $r_0 = 0$ and $r_{\mathrm{CTC}} < 0$.

For the KN spacetime, while the equatorial ring singularity is located at $r_0=0$, its causality-violating region extends to the positive$-r$ region, i.e., $r_{\mathrm{CTC}} > r_0$ (see Sec. 4.2 of Ref. \cite{Wuthrich1999}; See also Ref. \cite{Prasad+2018}). Note, however, that these regions are constrained to the interior of the KN inner horizon. 

For the KH regular BH spacetime, we find the largest root of $\Pi = 0$ to be located at $r=0$. However, immediately inside, $\Pi$ returns to being positive. Deeper inside the negative$-r$ region, we do find a causality-violating region. However, this appears at a radial coordinate smaller than that of a surface that has $\Pi \rightarrow \infty$ (i.e., large circumferential radius). This surface, therefore, corresponds to the inner asymptotic limit of the spacetime. We see, in this way, that the causality-violating region is causally-disconnected from the remaining regions we have discussed (e.g., from null infinity), and we discard this region as being a mathematical artifact. In summary, the KH regular BH spacetime is also devoid of causality-violating regions. 

For the KZM4 regular BH spacetime, $\Pi$ first vanishes at $r=0$ but remains positive until its second largest root at some negative $r$. Inside that surface, we find a causality-violating region of finite extent. As discussed above, we will set $r_{\mathrm{CTC}} = 0$ for this spacetime as well.

For the KS spacetime, we find $r_0 = r_{\mathrm{CTC}} = Q^2/M$ (note that $R(Q^2/M)=0$; Table \ref{table:Known_Static_Solutions}). 

Finally, for the KF spacetime, we find $r_0 = 0$. While $\Pi$ also first vanishes at this location, similar to the KH case, it becomes negative only well into the negative$-r$ region, which is causally-disconnected from null-infinity. Thus, the KF spacetime contains a curvature singularity but not a causality-violating region.

Therefore, for all barring the KN metric, we set $R(r_{\mathrm{CTC}}) = 0 \Rightarrow A(r_{\mathrm{CTC}}) = 0$, leading to the $\Pi=0$ surface being bounded inside an areal-radius of (see eq. \ref{eq:CTC_Area})
\begin{equation}
\mathscr{R}_{\mathscr{A}}(r_{\mathrm{CTC}}) = a/\sqrt{2}\,.
\end{equation}
The left panel of Fig. \ref{fig:FigA1_Areal_Radius_CTCs} in \ref{app:AppB_CTCs} shows the variation of the areal radius as a function of the coordinate radius in all spinning spacetimes considered here. The right panel there confirms that the regions of spacetime considered below are indeed CTC-free. In summary, by using the areal-radius, we automatically discount the causality-violating regions of these spacetimes.


\subsection{Black Hole Spacetimes with Rigidly Rotating Matter}
\label{sec:SecVB_Rigidly_Rotating}


\begin{figure}
\centering
\includegraphics[width=\columnwidth]{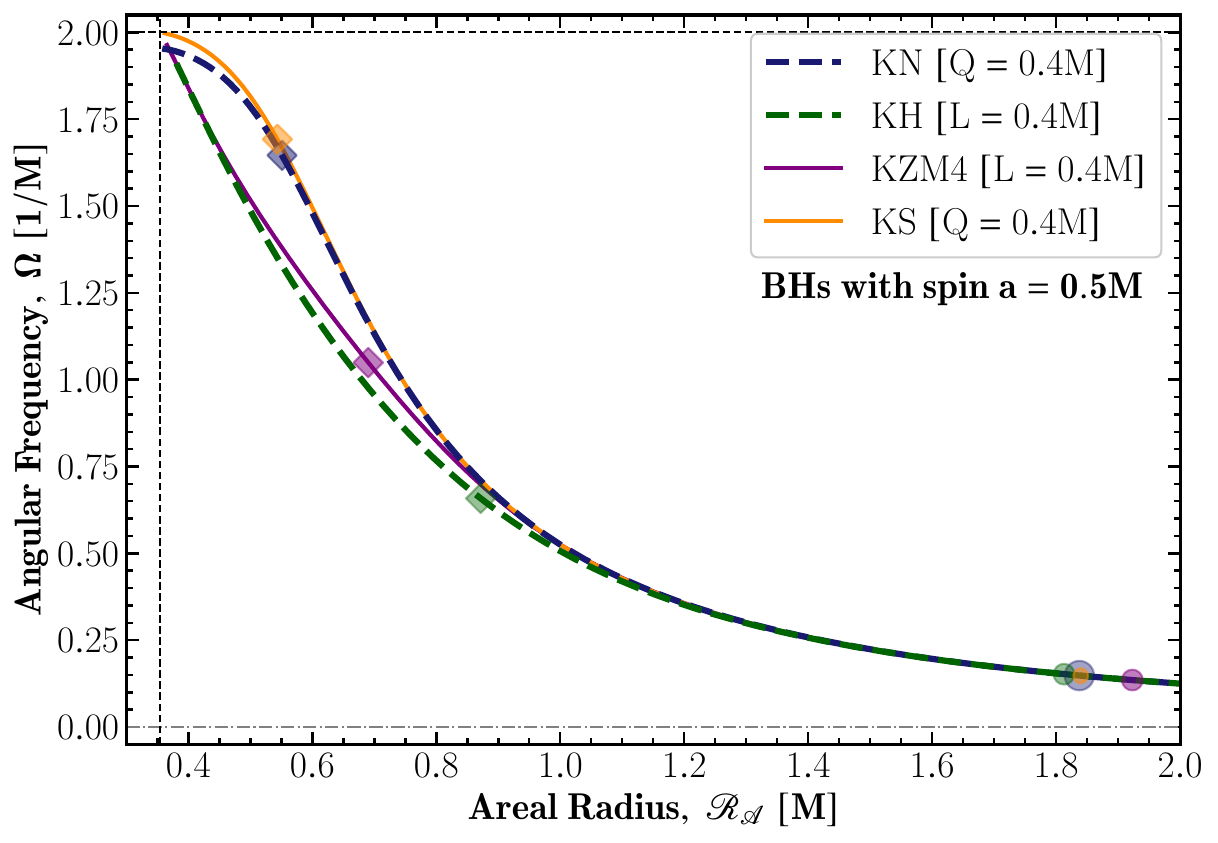}
\caption{Shown here is the angular velocity as a function of areal radius of the axially-spinning matter that generates various nonvacuum, axisymmetric black hole (BH) spacetimes. The angular velocity is constant over each Boyer-Lindquist $2-$sphere, indicating rigid rotation at each radius. The circles and diamonds locate the outer and inner horizons. The upper horizontal line, $\Omega = 1/a$ \eqref{eq:Omega_R0}, corresponds to the angular velocity of the surface $R(r)=0$ (vertical line). For both the KS and KF spacetimes, these correspond to the angular velocity and the location of the equatorial ring singularity respectively. For the KN spacetime, this surface lies inside the causality-violating region.}
\label{fig:Fig3_Omega_PNC}
\end{figure}

\begin{figure*}
\centering
\includegraphics[width=2\columnwidth]{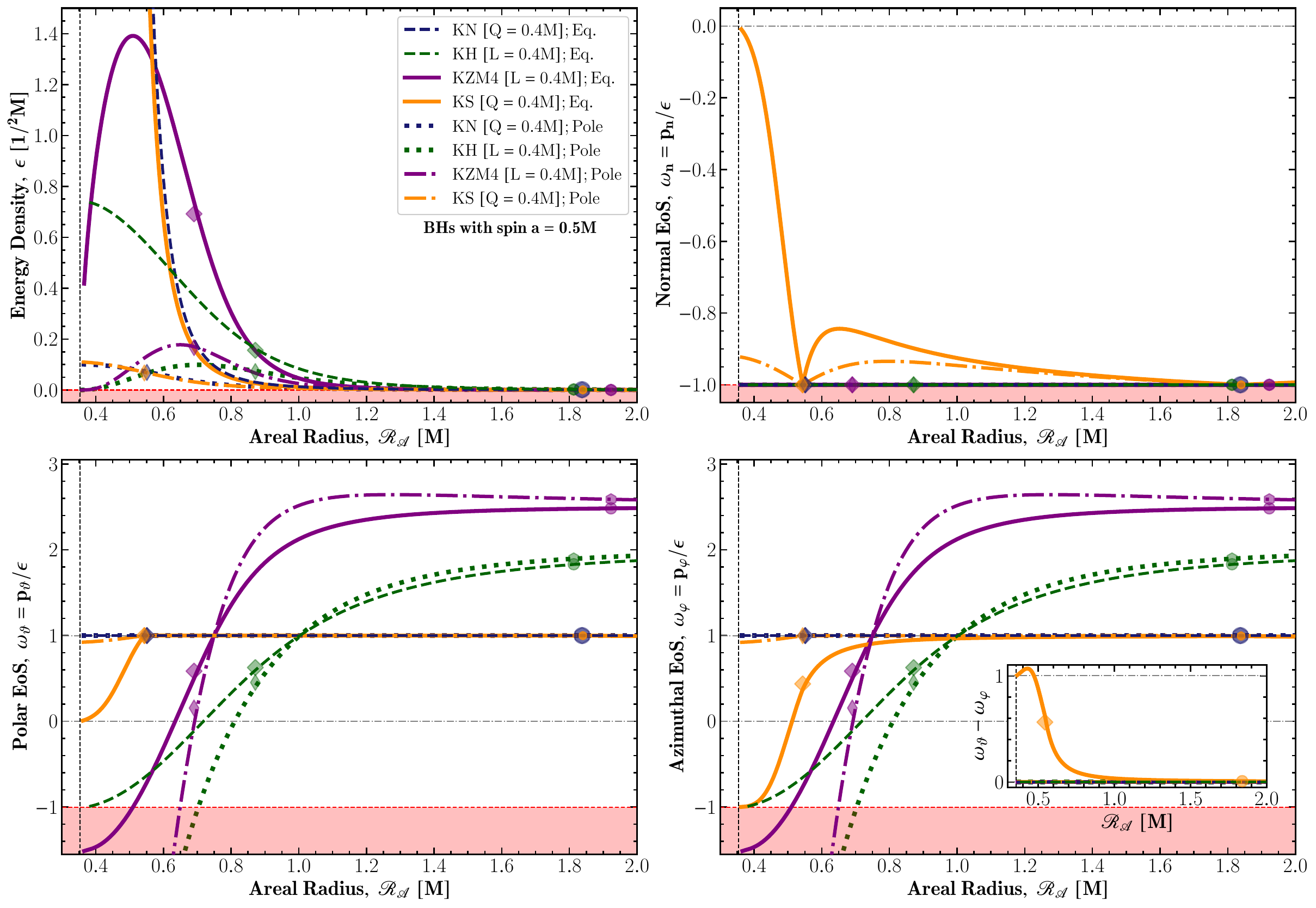}
\caption{The top left panel here shows the spatial distribution of the energy density of the background matter in its rest-frame in various spinning black hole (BH) spacetimes. The energy density is positive-definite everywhere. The circles and diamonds locate the outer and inner horizons. The rest-frame principal equations of state (EoSs) are shown in the remaining panels. The normal EoS is identically equal to $-1$ for the degenerate spacetimes (KN, KH, KZM4). Furthermore, the two tangential EoSs, $\omega_\vartheta$ and $\omega_\varphi$, are equal for these special spacetimes (see inset in bottom-right panel), even though the spacetime is not spherically-symmetric. For the KS metric, which is not a degenerate spacetime, the normal EoS becomes $-1$ at the two horizons, and the tangential EoSs do not match in the equatorial plane. We note that while the KN, KH, and KZM4 BH spacetimes contain a single component matter field, the KS metric contains a scalar (dilaton), an electromagnetic, and a pseudoscalar (axion) field. The singular BH spacetimes KN metric satisfies all the classical energy conditions (pure electromagnetic field). The regular BH spacetimes (KH, KZM4) violate them all (red region), albeit only well inside the inner horizon. The singular KS metric satisfies all energy conditions barring the strong energy condition, which is violated inside the inner horizon.}
\label{fig:Fig4_Matter_Distribution}
\end{figure*}


Fig. \ref{fig:Fig3_Omega_PNC} shows the matter angular velocity $\Omega$ for the KN, KH, KZM4, and KS BH spacetimes, as a function of the areal-radii $\mathscr{R}_{\mathscr{A}}(r)$ of $2-$spheres. Since all of these spacetimes satisfy the condition given in eq. \ref{eq:OmegaPNC_Condition}, the matter angular velocity matches that of the PNCs, $\Omega = \Omega_{\mathrm{PNC}}$. Since the latter is a function of $r$ alone, we find that the matter in these spacetimes is rigidly-rotating on any given coordinate $2-$sphere in the stationary regions of spacetime. The plot covers both the exterior and the interior regions of these BHs. Between the two horizons, $\Omega$ denotes the precession frequency of the matter rest-frame spatial axes, as discussed above. The vertical line corresponds to the surface $R(r) = 0$ for all barring the KN spacetime,
on which (see eq. \ref{eq:Omega_PNC}),
\begin{equation} \label{eq:Omega_R0}
\Omega = 1/a\,,
\end{equation}
as verified by the upper horizontal line. For the KS and KF spacetimes, this corresponds to the angular velocity of the equatorial timelike ring singularity. For the KN spacetime since $r=r_{\mathrm{CTC}} > 0$, $R=R(r_{\mathrm{CTC}}) > 0$ (Table \ref{table:Known_Static_Solutions}).

Since the matter angular velocity defines the matter rest-frame completely \eqref{eq:Comoving_Frame_Stationary_regionI}, we can project the Einstein tensor onto this tetrad to obtain the rest-frame energy density $\epsilon$ and principal pressures $p_i$ \eqref{eq:Rest_Frame_Matter_Properties}. The top-left panel of Fig. \ref{fig:Fig4_Matter_Distribution} shows $\epsilon$ as a function of the areal-radius in the equatorial plane as well as along the spin-axis in the four BH spacetimes discussed above. It is clear to see that the rest-frame energy density is nonnegative everywhere. It diverges in the equatorial plane for the KN and KS spacetimes, due to the presence of the curvature singularity, and remains regular everywhere for the regular BH spacetimes. For the KZM4 metric in particular, the energy density vanishes at $r=0$.

The remaining panels of Fig. \ref{fig:Fig4_Matter_Distribution} show the spatial-variation of the three equations of state (EoSs) $\omega_i = p_i/\epsilon$. The top-right panel shows that the normal EoS for all the \textit{degenerate} spacetimes (KN, KH, KZM4) is identically equal to $-1$. For the KS metric, which is a nondegenerate spacetime, the normal EoS shows nontrivial variation. Nonetheless, we see that it converges to $-1$ at both its horizons, as anticipated by eq. \ref{eq:Horizon_NEC}. This is remarkable since, even though the spacetime contains three different matter fields (including a \textit{massless scalar}), with nonminimal couplings between them (in the Einstein frame), their pressure distributions manage to conspire to produce a total normal EoS of $\omega_n = -1$ at the two horizons.  

Combining all the four panels of Fig. \ref{fig:Fig4_Matter_Distribution}, we see that the rest-frame energy-momentum-stress (EMS) tensor for the KN metric, which contains only an electromagnetic field, is precisely $\mathscr{T}_{(a)(b)} = \mathrm{diag.}[\epsilon, -\epsilon, \epsilon, \epsilon]$, as derivable from first-principles (see \ref{app:AppF2_EM_Fields}).

Furthermore, while the polar and azimuthal EoSs for the two regular BH spacetimes (KH, KZM4) display nontrivial variation with radius and polar angle, the inset makes clear that the two tangential EoSs are identically equal to each other everywhere in these degenerate spacetimes. This is striking due to the absence of spherical-symmetry in such spinning BH spacetimes. 


\begin{figure*}
\centering
\includegraphics[width=2\columnwidth]{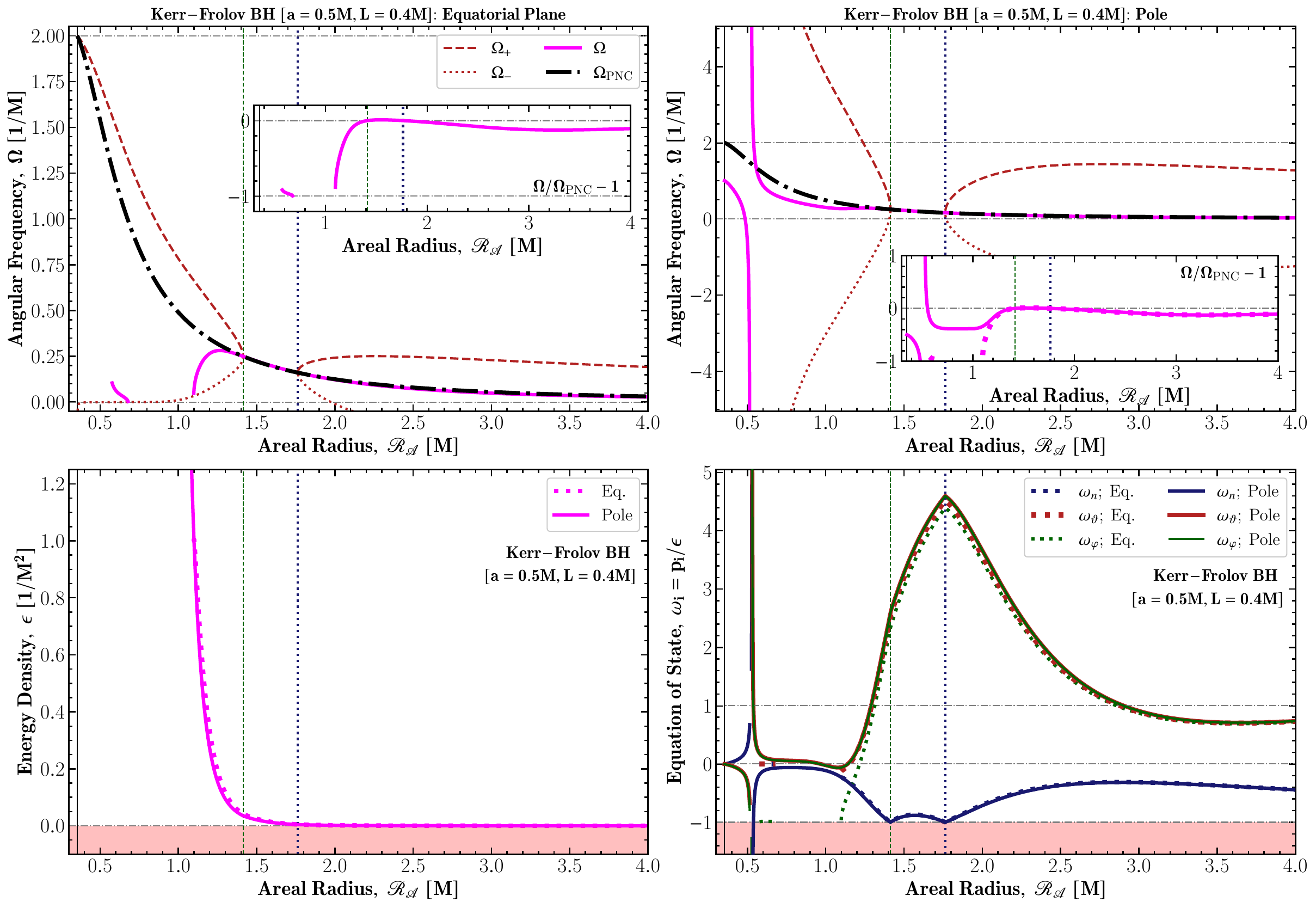}
\caption{The top-left panel shows the variation of the matter angular velocity $\Omega$, as a function of the $2-$sphere areal-radius $\mathscr{R}_{\mathrm{A}}$, in a Kerr-Frolov (KF) black hole (BH) spacetime in the equatorial plane, and the top-right panel shows the same along the BH spin axis. The solid vertical line locates the equatorial ring singularity. The matter angular velocity is obtained from eq. \ref{eq:Omega_Quadratic}, and lies between the critical null angular velocities $\Omega_\pm$, ensuring the matter flows on timelike orbits. The gaps in $\Omega$ in the region $\mathscr{R}_{\mathrm{A}} \lesssim 1.1M$, i.e., inside the inner horizon (vertical dashed line), indicate that eq. \ref{eq:Omega_Quadratic} returns complex roots for $\Omega$. The two panels also show the matter to be differentially-rotating on each coordinate $2-$sphere, and the insets reveal that $\Omega \neq \Omega_{\mathrm{PNC}}$ \eqref{eq:Omega_PNC}, except at the horizons (the dotted line locates the event horizon). The bottom-left panel shows the rest-frame energy density of the background matter to be positive-definite throughout the region $\mathscr{R}_{\mathrm{A}} \gtrsim 1.1M$, and the bottom-right panel shows the principal equations-of-state (EoSs; eq. \ref{eq:EoS}). Together, these show that the KF matter obeys all energy conditions barring the dominant energy condition for $\mathscr{R}_{\mathrm{A}} \gtrsim 1.1M$. The normal EoS takes its expected value, $\omega_n = -1$, everywhere on a horizon (Sec. \ref{sec:SecIIIA_Horizons}).}
\label{fig:Fig5_Kerr_Frolov_Master}
\end{figure*}

\begin{figure*}
\centering
\includegraphics[width=2\columnwidth]{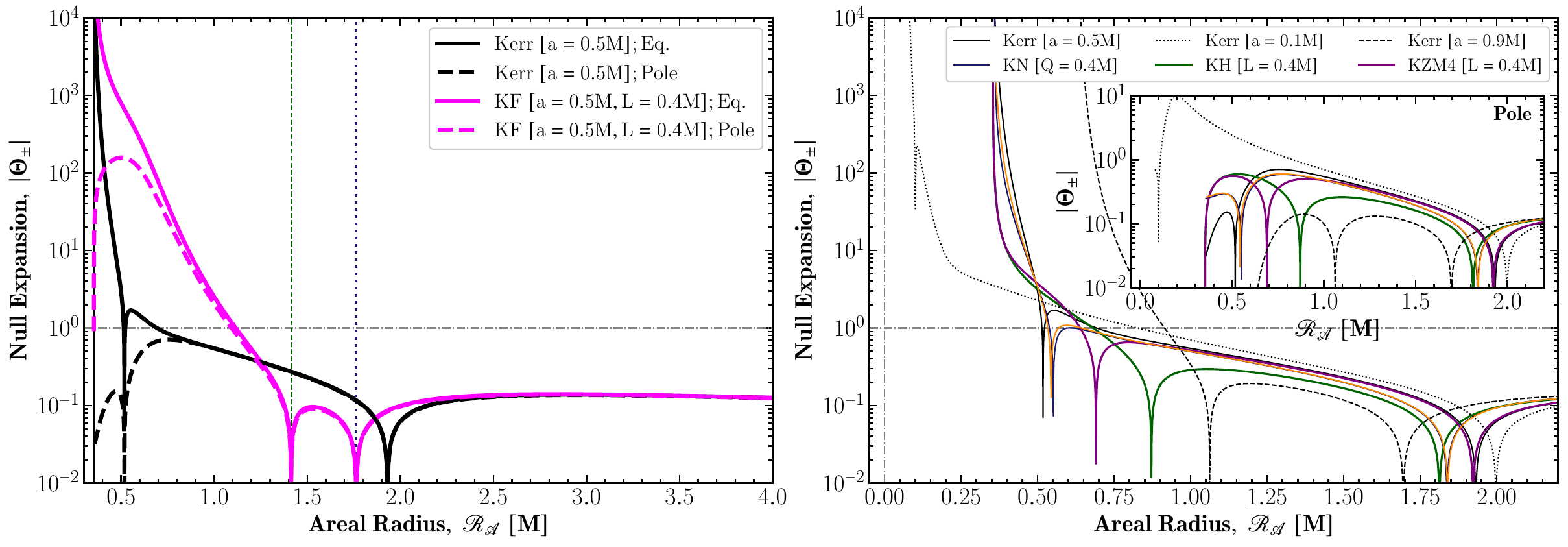}
\caption{
Shown here are the absolute values of the expansions of two fundamental zero angular momentum null congruences, which measure the strength of light-focusing, as a function of the areal-radius of Boyer-Lindquist coordinate $2-$spheres, in various black hole (BH) spacetimes. These correspond to the fractional rate of change of the cross-sectional area of the congruences along the congruence \eqref{eq:PNEs}. While the ingoing ($-$) null expansion is always negative (implying convergence), the outgoing ($+$) one is positive outside the event horizon and inside the inner horizon but is negative between the two horizons (see Sec. \ref{sec:SecIID_PNCs}). Both expansions vanish at the horizons. In the left-panel, we show the expansions in the Kerr and Kerr-Frolov BH spacetimes, both in the equatorial plane as well as along the BH spin axis. It is clear that the null expansions vary in magnitude over the surface of a $2-$sphere. The right panel similarly shows the expansions for the remaining spinning BH spacetimes considered here in the equatorial plane whereas the inset shows the same along the pole, to demonstrate the genericity of the nonmonotonicity of the null expansions between the horizons, i.e., in the contracting BH interior cosmology.}
\label{fig:Fig6_PNEs}
\end{figure*}


The KS metric again displays nontrivial variation of the tangential EoS with $r$ and $\theta$, but with the additional feature that the two tangential EoSs, $\omega_\vartheta$ and $\omega_\varphi$, differ significantly from each other in the equatorial plane (inset).

With the nonnegativity of the energy density and the spatial variation of the principal EoSs for each of these spacetimes in hand, we can finally comment on the validity of the classical energy conditions in these spacetimes. 

The KN BH spacetime satisfies all energy conditions (see \ref{app:AppD_Energy_Conditions}), as can be seen from its rest-frame EMS tensor above. 

Since the tangential EoSs for the two regular degenerate BH spacetimes enter the red regions in the bottom panels ($\omega_\vartheta = \omega_\varphi < -1$, or, equivalently, $p_\vartheta + \epsilon = p_\varphi + \epsilon < 0$), it becomes clear that they violate the null energy condition (NEC), albeit deep inside the inner horizon (see also Ref. \cite{Bambi+2013}). Therefore, they also violate the weak energy condition (WEC), the dominant energy condition (DEC), as well as the strong energy condition (SEC). It is interesting to note that the KH metric does not violate the NEC or the WEC in the equatorial plane anywhere, underscoring the importance of carefully examining the properties of matter throughout the entire spacetime. In the BH exterior, however, the NEC, WEC, and SEC are satisfied but the DEC remains violated ($\omega_\vartheta = \omega_\varphi > 1$).

Finally, the KS spacetime satisfies the NEC, WEC, and DEC everywhere. However, the SEC is violated inside the inner horizon. The latter is seen from the value of the EoSs at equatorial singularity, $\omega_r < 0, \omega_\vartheta = 0,$ and $\omega_\varphi = -1$, implying $\omega_r + \omega_\vartheta + \omega_\varphi < -1$. 


\subsection{The Kerr-Frolov Black Hole Spacetime}
\label{sec:SecVC_Non_Rigidly_Rotating}

In the previous section, we considered a variety of BH models whose matter was seen to be rigidly-rotating over each coordinate $2-$sphere. In this section, we discuss the properties of the novel Kerr-Frolov (KF) BH spacetime, for which this is not the case. The KF model is different from the other models because we do not have an analytical spherically symmetric seed metric with $g(r)=1$. We therefore retain the non-trivial $g(r)$, which propagates into the spinning metric (see $\mathscr{g}_{rr}$) via our ansatz equation \ref{eq:Stationary_Metric}. An immediate consequence is that we no longer have $\Omega = \Omega_{\rm PNC}$. In addition, as we describe below, $\Omega$ at a given $r$ now varies with $\theta$.

In the top row of Fig. \ref{fig:Fig5_Kerr_Frolov_Master}, the solid line shows the angular velocity of the background matter, $\Omega$, in a non-extremal KF BH spacetime (with $a=0.5M$ and $L=0.4M$), in the equatorial plane (top left panel) and on the spin axis (top right panel), as a function of the areal radius $\mathscr{R}_{\mathrm{A}}(r)$ of a $2-$sphere of coordinate radius $r$. The solid vertical line in all panels shows the location of the curvature ring singularity. The rest-frame angular velocity is obtained by solving the quadratic equation eq. \eqref{eq:Omega_Quadratic} for a real root that lies between the critical null angular velocities $\Omega_+$ (dashed line) and $\Omega_-$ (dotted). As a reminder, the condition, $\Omega_- < \Omega < \Omega_+$ ensures that the matter $4-$velocity is timelike \eqref{eq:Killing_Norm}. 

Interestingly, we find regions inside the inner horizon (vertical dashed line), where no real solution to eq. \ref{eq:Omega_Quadratic} is obtained. We interpret this to mean that the matter flows on achronal orbits in such regions.%
\footnote{We also note that $\mathscr{g}_{\varphi\varphi}$ is positive-definite everywhere outside the ring singularity $r > 0$ (\ref{app:AppB_CTCs}).} %
The structure of these regions is nontrivial, as can be seen by comparing the top-left and top-right panels. We see that the KF matter flows on timelike orbits everywhere outside an areal-radius of $\mathscr{R}_{\mathrm{A}} \approx 1.1M$, corresponding, equivalently, to a BL coordinate radius of $r \approx 1M$. Inside this radius, the metric is pathological.

Such comparison also reveals that the matter is differentially-rotating, i.e., $\Omega = \Omega(r, \vartheta)$, on each coordinate $2-$sphere, except at the event horizon (vertical dotted line) and the inner horizon, where the matter angular velocity matches that of the horizon. We note that, although not visible here, outside the inner-horizon, the angular velocity at the equator differs from that on the spin-axis by less than $0.1\%$.

In all of the BH models considered previously, we saw that the matter was not only rigidly-rotating on each coordinate $2-$sphere, but that it was rotating at a special frequency, namely that of the principal null congruences (PNCs), $\Omega_{\mathrm{PNC}}(r) = a/(A(r)+a^2)$. The dot-dashed lines in the top row panels in Fig.~\ref{fig:Fig5_Kerr_Frolov_Master} show $\Omega_{\mathrm{PNC}}$ for the KF metric. The insets in each panel confirm that we have a nonzero (but mostly moderate, $\sim 10\%$) fractional deviation of the matter angular velocity $\Omega$ from $\Omega_{\mathrm{PNC}}$ (except at the horizons). 

The bottom-left panel in Fig. \ref{fig:Fig5_Kerr_Frolov_Master} shows the matter rest-frame energy density profile and the bottom-right panel shows the principal equations of state. In the region $\mathscr{R}_{\mathrm{A}} \gtrsim 1.1M$, we find the energy density to be positive-definite, and all classical energy conditions, except for the dominant energy condition, to be satisfied. This includes the nonstationary region between the two horizons, which is interesting because the matter in the nonspinning Frolov metric contained energy-condition-violating regions between the horizons. 

For completeness, we note that as we move further inwards along the pole, approaching the surface where $\Omega$ first ceases to be real-valued, $\mathscr{R}_{\mathrm{A}} \approx 0.5M$, we find the energy density to take negative values (not shown here due to scale). Furthermore, the null energy condition is clearly violated close by since $\omega_r < -1$ (bottom-right panel). 


\subsection{ZAM Null Expansions \& The Interior}
\label{sec:SecVD_PNEs_Results}

As discussed above (Sec. \ref{sec:SecIID_PNCs}), the zero angular momentum null congruences capture the global causal structure of the spacetime. Their expansions in particular can be quite useful in visualising the nontrivial BH interior cosmology.

Fig. \ref{fig:Fig6_PNEs} shows the absolute values of the null expansions in all the spinning spacetimes covered here, both in the equatorial plane as well as along the black hole spin axis. The mismatch in the magnitudes of the expansions across colatitudes (different $\vartheta$) shows differential rates of expansion over any particular coordinate $2-$sphere. 

We emphasize that each component of the two null congruence generators differs only in sign and not in magnitude both in the stationary \eqref{eq:NP_Null_Tetrad_RegionI} as well as the nonstationary \eqref{eq:NP_Null_Tetrad_RegionII} regions of spacetime. This enables introducing a gauge-invariant definition for each individual null expansion as $|\Theta_\pm| = \sqrt{|\Theta_+\Theta_-|}$. As described in Sec. \ref{sec:SecIID_PNCs}, the ingoing null expansion is always negative whereas the outgoing one is positive in the stationary regions of spacetime but negative in the nonstationary regions of spacetime. Both null expansions vanish at a horizon since both vector fields limit to the horizon generators.  

It follows from the above then that the two expansions must display nonmonotonic behavior in the interior cosmology, i.e., between the two horizons. This is evident to see from the figure. This suggests that while the cross-sectional area monotonically decreases (both expansions are negative, $\Theta_\pm < 0$), the evolution of the expansion scalar switches character. We understand the latter by remembering that the radial coordinate $r$ can be used to parameterize these congruences (see \ref{app:AppC_Null_Kerr_Schild}), and considering the derivative of the null expansions with respect to it. By analogy with the deceleration parameter in standard cosmology, this suggests that the BH interior cosmology is initially accelerating upon crossing the event horizon but then switches to being decelerating before reaching the inner horizon. More detailed insight into this aspect will require a careful analysis of the Raychaudhuri equation \cite{Raychaudhuri1953}, which we do not attempt here. Similarly, classifying these cosmologies (see, e.g., Sec. 11.3.3 and 11.3.4 of Ref. \cite{Hawking+2010}) is beyond the scope of the present article.


\section{Summary \& Conclusions}
\label{sec:SecVI_Conclusions}

Given a general nonspinning and spherically-symmetric metric of the form
\begin{equation} \label{eq:Nonspinning_Metric}
\mathrm{d}s^2 
= -\mathrm{e}^{2\psi}\left(1-\frac{2m}{R}\right)\mathrm{d}t^2 + \frac{(\partial_r R)^2}{\left(1-\frac{2m}{R}\right)}~\mathrm{d}r^2 + R^2~\mathrm{d}\Omega_2^2\,,
\end{equation}
where the metric functions $R(r)$ and $m(r)$ denote the areal-radius of a $2-$sphere of coordinate radius $r$ and its Hawking mass respectively, and $\psi(r)$ is the redshift function that measures aspects of the pressure profile of the background matter, the Azreg-A{\"i}nou (AA) solution-generating algorithm \cite{Azreg-Ainou2014a, Azreg-Ainou2014b, Azreg-Ainou2014c, Azreg-Ainou2015} proposes an associated class of unique spinning and axisymmetric generalization metrics, whose form is given as
\begin{align} \label{eq:Stationary_Metric_v2}
\mathrm{d}s^2
=&\ 
-\left(1-\frac{2 F}{\Sigma}\right)\mathrm{d}t^2 
-2\frac{2 F}{\Sigma}a\sin^2{\vartheta}~\mathrm{d}t\mathrm{d}\varphi
+\frac{\Pi}{\Sigma}\sin^2{\vartheta}~\mathrm{d}\varphi^2 \nonumber \\ 
&\ 
+ \frac{\Sigma}{\Delta}g~\mathrm{d}r^2 + \Sigma~\mathrm{d}\vartheta^2\,.
\end{align}
The function $g$ above is a product of the $tt-$ and $rr-$components of the nonspinning metric \eqref{eq:Nonspinning_Metric}, 
\begin{equation}
g(r) = \left(\mathrm{e}^\psi\cdot\partial_rR\right)^2\,,
\end{equation}
and plays a crucial role in determining whether the spinning metric \eqref{eq:Stationary_Metric_v2} limits to the nonspinning one \eqref{eq:Nonspinning_Metric} conformally \cite{Azreg-Ainou2014a, Azreg-Ainou2014b, Azreg-Ainou2014c} or exactly \cite{Azreg-Ainou2015, Chen2022a} in the AA framework (see Sec. \ref{sec:SecII_Metric_Ansatz} and \ref{app:AppG_AA_Comparison}).

In the above, by uniqueness, we simply mean that the spinning metric functions $F, \Delta, \Sigma$, and $\Pi$ are completely fixed by the nonspinning ones:
\begin{equation}
\begin{alignedat}{3}
&\ F(r) = (A - B)/2\,;\ 
&&\ \Delta(r) = B + a^2\,, \\
&\ \Sigma(r, \vartheta) = A + a^2\cos^2{\vartheta}\,;\ 
&&\ \Pi(r, \vartheta) = \left(A + a^2\right)^2 - \Delta a^2\sin^2{\vartheta}\,,
\end{alignedat}  
\end{equation}
with 
\begin{align}
A(r) = R^2\,;\ \ B(r) = \mathrm{e}^{2\psi}(R^2 - 2mR)\,.
\end{align}
The axisymmetric metric \eqref{eq:Stationary_Metric_v2} is asymptotically-flat, permits a Carter constant for all geodesics, and can be used to describe several popular black hole (BH) spacetimes (Sec. \ref{sec:SecV_Spinning_BH_Properties}). Since we are interested in studying the properties of the spacetime-generating, self-gravitating matter in a broad class of stationary black hole (BH) spacetimes in this article, both in the BH exterior as well as in its interior, the algebraic simplicity of this metric facilitates great physical clarity, for our purposes.

In general, the metric \eqref{eq:Stationary_Metric_v2} possesses a nontrivial Einstein tensor, indicating that it describes a nonempty spacetime. We describe a simple method to identify the timelike $4-$velocity of the axially-spinning background matter, 
\begin{equation} \label{eq:Matter_4_Vel}
e_{(t)}^\mu \propto [1, 0, 0, \Omega]\,,
\end{equation}
which involves solving a single quadratic equation \eqref{eq:Omega_Quadratic} for a real-valued angular frequency $\Omega$. This allows us to unambiguously introduce the rest-frame or comoving-frame of the matter distribution, both in the BH exterior as well as in its interior (Sec. \ref{sec:SecIII_Comoving_Frame}). We then obtain the rest-frame energy density $\epsilon$ and principal pressures $p_i$ of the matter by projecting the Einstein tensor onto the matter rest-frame \eqref{eq:Rest_Frame_Matter_Properties}. 

With the above, it is found that at a BH horizon of the axisymmetric metric \eqref{eq:Stationary_Metric_v2}, the matter rest-frame energy density and normal pressure $p_n$ always satisfy,
\begin{equation} \label{eq:pr_plus_rho_Eq}
p_n = -\epsilon\,.
\end{equation}
We highlight that the equation above yields powerful ``no-go'' theorems for the type of permissible matter content in stationary BH spacetimes. For example, for a massless-scalar field minimally-coupled to Einstein-Hilbert gravity in a spherically-symmetric configuration, $p_n = \epsilon$ in the matter rest-frame (\ref{app:AppF1_Scalar_Fields}). Therefore, such matter by itself cannot thread either a static or a stationary BH horion (cf. also Ref. \cite{Bekenstein1972}). This has interesting consequences for the solution space of general relativity minimally-coupled to matter, as well as for the cosmic censorship hypothesis \cite{Penrose2002}.

Spherically-symmetric spacetimes filled with matter content that satisfies \eqref{eq:pr_plus_rho_Eq} \textit{everywhere} are special. Such spacetimes are described by a single metric function, i.e., their Hawking mass function $m(r)$ (with $\psi = 0$ and $R = r$). Various interesting properties of these ``degenerate'' spacetimes have been discussed in Ref. \cite{Jacobson2007}. In Sec. \ref{sec:SecIIIB_Degenerate_Spacetimes}, we review how degenerate regular BH spacetimes (no curvature singularity at the center; cf. Ref. \cite{Zhou+2023}), typically violate all energy conditions (cf. also Ref. \cite{Bambi+2013}), close to the center $r=0$ \eqref{eq:Center_EoS_Regular}. 

In \ref{app:AppB_CTCs} and Sec. \ref{sec:SecIID_PNCs}, we obtain simple analytic expressions for the area of a coordinate $2-$sphere \eqref{eq:Area} and for the (\textit{gauge-invariant}) expansions of the fundamental zero angular momentum null congruences of the spacetime \eqref{eq:Null_Expansion_RegI} respectively, both in the BH exterior as well as in its interior. The former allows us to introduce the areal-radius, which enables meaningful comparisons of various spatially-varying quantities across different spacetimes. The latter plays a vital role in revealing the causal structure of the spacetime.

In Sec. \ref{sec:SecIV_Nonspinning_BH_Properties}, we apply the general framework we have developed in the previous sections to examine the properties of matter and spacetime in six repesentative nonspinning BH spacetimes (Table \ref{table:Known_Static_Solutions}). Of these, four are degenerate BH spacetimes. The first of these is the Schwarzschild BH metric \cite{Schwarzschild1916}, which describes a vacuum spacetime. We review how the Reissner-Nordstr{\"o}m (RN) solution (see, e.g., Ref. \cite{Wald1984}) satisfies all classical energy conditions. This is to be expected since the spacetime contains an electromagnetic field (see \ref{app:AppF2_EM_Fields}). The Hayward regular BH metric \cite{Hayward2006} is an interesting exception to the aforementioned rule for degenerate regular BH spacetimes. Its matter content only violates the dominant energy condition (DEC) since its rest-frame tangential pressure $p_\vartheta$ exceeds its rest-frame energy density in magnitude, i.e., $p_\vartheta > \epsilon$. The Zhou-Modesto-4 (ZM4) regular BH metric \cite{Zhou+2023} violates all energy conditions but only inside its inner horizon. Outside the inner horizon, it only violates the DEC ($p_\vartheta > \epsilon$). 

We also re-examine the matter content in the Gibbons-Maeda-Garfinkle-Horowitz-Strominger (GMGHS) BH metric \cite{Gibbons+1988, Garfinkle+1991}, which is a solution of a particular Einstein-Maxwell-dilaton/scalar theory. Since the GMHGS spacetime does not belong to the class of degenerate spacetimes (Table \ref{table:Known_Static_Solutions}), we find a nontrivial normal pressure profile, i.e., $p_n \neq - \epsilon$ everywhere. However, precisely at the horizon, we do find $p_n=-\epsilon$, as required by equation \ref{eq:pr_plus_rho_Eq}. As an aside we note that this BH does not possess an inner (Cauchy) horizon, and that its surface gravity as well as its causal structure are identical to that of a Schwarzschild BH. We find that its matter fields combined satisfy all classical energy conditions. 

Finally, we consider the Frolov regular BH metric \cite{Frolov2016}, which was proposed as a novel modification of the Hayward metric. While its Hawking mass function $m$ is identical to that of the Hayward metric, its redshift function $\psi$ is nontrivial. We find that the matter content violates all energy conditions well inside the event horizon. In the exterior horizon geometry, only the DEC ($p_\vartheta > \epsilon$) is violated. In each of these BH spacetimes, the matter content obeys eq. \ref{eq:pr_plus_rho_Eq} at every horizon (Fig. \ref{fig:Fig2_Energy_Density_Pressures_ECs}).

In Sec. \ref{sec:SecV_Spinning_BH_Properties}, we consider the properties of matter and spacetime in the spinning counterparts of the nonspinning spacetimes discussed in Sec. \ref{sec:SecIV_Nonspinning_BH_Properties}, as described by the metric \eqref{eq:Stationary_Metric_v2}. More specifically, the zero-spin ($a=0$) limits of the six BH metrics, Kerr \cite{Kerr1963}, Kerr-Newman (KN; \cite{Newman+1965b}), Kerr-Hayward (KH; \cite{Bambi+2013}), Kerr-ZM4 (KZM4), Kerr-Sen (KS; \cite{Sen1992}), and Kerr-Frolov (KF), correspond to the nonspinning BH metrics described in Sec. \ref{sec:SecIV_Nonspinning_BH_Properties} and listed in Table \ref{table:Known_Static_Solutions}. Strikingly, we find that while the nonspinning Frolov BH is everywhere regular, the KF BH picks up a curvature singularity in this solution-generating technique.

In all of these nonvacuum spinning spacetimes, the self-gravitating matter flows on circular orbits about the BH spin axis \eqref{eq:Matter_4_Vel} in the stationary regions of spacetime (i.e., outside the event horizon and inside the inner horizon; See Sec. \ref{sec:SecIIA_Ansatz_Metric_Properties}). In the nonstationary-region of spacetime (i.e., between the two horizons), the matter comoves with the interior cosmology \eqref{eq:Comoving_Frame_Stationary_regionII}, i.e., it exhibits no spatial motion in the interior Boyer-Lindquist coordinates introduced in Sec. \ref{sec:SecIIC_Interior_Coordinates}.

Furthermore, in the spinning counterparts of degenerate nonspinning spacetimes specifically, we find the matter angular velocity $\Omega$ to be constant over each coordinate $2-$sphere in the stationary regions of spacetime. Additionally, it is given by a special value there (see Refs. \cite{Azreg-Ainou2014a, Azreg-Ainou2014b, Azreg-Ainou2014c}),
\begin{equation} \label{eq:RF_Degen}
\Omega(r) = \frac{a}{R^2(r) + a^2} = \Omega_{\mathrm{PNC}}(r)\,.
\end{equation}
This is precisely the angular velocity of both the ingoing and the outgoing PNCs, $\Omega_{\mathrm{PNC}}$ \eqref{eq:Omega_PNC}. Therefore, of the six spacetimes considered here, the background matter in the stationary regions of the KN, KH, and KZM4 spacetimes is rigidly-rotating on each coordinate $2-$sphere. This property is also found to hold for the KS metric (see \ref{app:AppG_AA_Comparison}), consistent with the prediction in Sec. 3 of Ref. \cite{Azreg-Ainou2014a}. 


\begin{table}[!ht]
\begin{center}
\caption{\textit{Summary of salient properties of matter in various black hole (BH) spacetimes.} Listed alternately below are the various nonspinning BH models and their spinning counterparts considered here. We highlight the similarities and differences between the different spacetimes by showing two of their metric functions, $R(r)$ and $g(r)$. The matter rest frame angular velocity $\Omega$ in each spacetime is obtained from a quadratic equation \eqref{eq:Omega_Quadratic}. For the degenerate BH spacetimes (Sec. \ref{sec:SecIIIB_Degenerate_Spacetimes}) as well as for the KS metric, the matter is rigidly-rotating on each coordinate $2-$sphere \eqref{eq:RF_Degen}. It is also clear that the two tangential equations of state, $\omega_\vartheta$ and $\omega_\varphi$, are unequal only in the two non-degenerate spacetimes. Finally, we indicate whether the weak energy condition (WEC), the dominant energy condition (DEC), and the strong energy condition (SEC) are satisfied ($\checkmark$) or violated ($\times$). In the nonpathological regions of these spacetimes, the energy-density is non-negative. Entries denoted by $\mathrm{RI}\checkmark$ denote that the energy condition is satisfied in the BH exterior (region I) but not in the interior.}
\label{table:Matter_Prop}
\renewcommand{\arraystretch}{1.5}
\centering
\begin{tabular}[t]{|| l || l | l || l || l || l | l | l ||}
\hhline{||========||}
Metric & $R(r)$ & $g(r)$ & $\Omega$ & $\omega_\varphi$ & WEC & DEC & SEC \\
\hhline{||========||}
Schw. & $r$ & $1$ & $-$ & $-$ & $-$ & $-$ & $-$ \\ 
Kerr & & & $-$ & $-$ & $-$ & $-$ & $-$ \\ 
\hline
RN & $r$ & $1$ & $0$ & $\omega_\vartheta$ & $\checkmark$ 
& $\checkmark$ & $\checkmark$ \\
KN & & & $\Omega_{\mathrm{PNC}}(r)$ & $\omega_\vartheta$ 
& $\checkmark$ & $\checkmark$ & $\checkmark$ \\
\hline
Hayward & $r$ & $1$ & $0$ & $\omega_\vartheta$ 
& $\checkmark$ & $\times$ & $\mathrm{RI}\checkmark$ \\
KH & & & $\Omega_{\mathrm{PNC}}(r)$ & $\omega_\vartheta$ 
& $\mathrm{RI}\checkmark$ & $\times$ & $\mathrm{RI}\checkmark$ \\
\hline
ZM4 & $r$ & $1$ & $0$ & $\omega_\vartheta$ 
& $\mathrm{RI}\checkmark$ & $\times$ & $\mathrm{RI}\checkmark$ \\
KZM4 & & & $\Omega_{\mathrm{PNC}}(r)$ & $\omega_\vartheta$
& $\mathrm{RI}\checkmark$ & $\times$ & $\mathrm{RI}\checkmark$ \\
\hhline{||========||}
GMGHS & $\neq r$ & $1$ & $0$ & $\omega_\vartheta$ 
& $\checkmark$ & $\checkmark$ & $\checkmark$ \\
KS & & & $\Omega_{\mathrm{PNC}}(r)$ & $\neq\omega_\vartheta$ 
& $\checkmark$ & $\checkmark$ & $\mathrm{RI}\checkmark$ \\
\hhline{||========||}
Frolov & $r$ & $\neq 1$ & $0$ & $\omega_\vartheta$ 
& $\mathrm{RI}\checkmark$ & $\times$ & $\mathrm{RI}\checkmark$ \\
KF & & & $\Omega(r, \vartheta)$ & $\neq\omega_\vartheta$ 
& $\mathrm{RI}\checkmark$ & $\times$ & $\mathrm{RI}\checkmark$ \\
\hhline{||========||}
\end{tabular}
\end{center}
\end{table}


For the KF metric alone we find the matter to be differentially-rotating on each coordinate $2-$sphere outside the BH, i.e., $\Omega = \Omega(r, \vartheta)$. At the event horizon itself, the matter rotates rigidly at the horizon angular velocity. This is similarly true for the matter at the inner horizon. We note, however, that the angular velocity in the equatorial plane differs from that on the spin axis by $\lesssim 0.1\%$ everywhere outside the inner horizon. Interestingly, inside the inner horizon, we find regions where the quadratic equation \eqref{eq:Omega_Quadratic} for $\Omega$ returns complex values. We interpret this as implying that, in such regions, the background matter flows on achronal orbits. In this way, the framework developed in this paper introduces a new ``sanity check'' on the physical reasonability of self-gravitating matter in spinning spacetimes.

We find the rest-frame energy-density of the matter to be positive-definite everywhere in the physically-reasonable regions of each of the spinning BH spacetimes we consider (i.e., when the matter flows on timelike orbits and in regions containing no closed timelike curves). The pure electromagnetic field in the singular RN and KN spacetimes is the only matter that satisfies all of the energy conditions in such regions. The regular BH spacetimes (Hayward, KH, ZM4, KZM4) and all violate the dominant energy condition (DEC). However, they satisfy the null energy condition (NEC), weak energy condition (WEC), and strong energy condition (SEC) in the BH exterior. The KS metric satisfies the NEC, WEC, and DEC outside the ring singularity but the strong energy condition (SEC) is violated inside the inner horizon. Its nonspinning limit, the GMGHS BH spacetime, satisfies all of the energy conditions everywhere. Finally, the matter in the KF BH metric satisfies the NEC, WEC, and SEC everywhere outside the event horizon. A summary of the important similarities and differences between these BH spacetimes and their matter content is presented in Table \ref{table:Matter_Prop} for easy access. Interestingly, the regular KH BH and the singular KF BH spacetimes do not contain causality-violating regions.

The spatial-profile, i.e., as a function of the areal-radius of coordinate $2-$spheres, of the expansions of the zero angular momentum null congruences generically exhibit nonmonotonic behavior behind the event horizon. In the interior cosmology in particular, this suggests that the cross-sectional areas of the principal null congruences decrease monotonically when approaching the center but while exhibiting accelerating and decelerating area-evolutions.

The analysis of the KZM4 and the KF BH spacetimes here demonstrates how when using the metric \eqref{eq:Stationary_Metric} as a ``solution-generating metric'' (i.e., to find a spinning, axisymmetric generalization for a given nonspinning, spherically-symmetric metric), the theoretical framework outlined here can be employed to systematically investigate the material properties of the self-gravitating matter content of the newly-generated spinning spacetime. While we restricted our attention to a specific class of axisymmetric spacetimes \eqref{eq:Stationary_Metric_v2}, we expect that the general principles underlying this framework can be extended to cover more general classes of axisymmetric spacetimes (e.g., Ref. \cite{Johannsen2013, Konoplya+2016}). 

As a final remark, we note that for non-fluid matter models in general, additional equations of motion for the matter field must also be solved to obtain a consistent solution to both the gravitational as well as the matter field equations. While we have not attempted this here, we direct the reader to Ref. \cite{Erbin2017} for the status and successes of the original Newman-Janis solution-generating technique.


\subsection*{Acknowledgments}
We thank Rahul Kumar Walia for stimulating discussions on black hole interiors. We thank Mustapha Azreg-A{\"i}nou, Naresh Dadhich, Madhusudhan Raman, Ronak Soni, and the referees for useful comments. Support comes from grants from the Gordon and Betty Moore Foundation (GBMF-8273) and the John Templeton Foundation (\#62286) to the Black Hole Initiative at Harvard University. 


\setcounter{section}{1}
\appendix


\section{The Azreg-A{\"i}nou Metric Solution Generating Algorithm}
\label{app:AppA_AA_Framework}

Shortly after the discovery of the Kerr metric, Newman \& Janis \cite{Newman+1965} proposed a novel algorithm, involving a complex coordinate transformation, that transforms the Schwarzschild metric \cite{Schwarzschild1916}, which describes nonspinning vacuum BHs, to the Kerr metric. Almost immediately thereafter, this algorithm was applied to the Reissner-Nordstr{\"o}m metric (see, e.g., Ref. \cite{Misner+1964}), a static and spherically-symmetric solution of the Einstein-Maxwell equations and used to describe nonspinning electrovacuum BHs, to successfully obtain the Kerr-Newman metric \cite{Newman+1965b}, a legitimate stationary and axisymmetric solution to the Einstein-Maxwell system that is used to describe spinning electrovacuum BHs in GR. This algorithm \cite{Yazadjiev2000} also converts the Gibbons-Maeda-Garfinkle-Horowitz-Strominger (GMGHS) metric \cite{Gibbons+1988, Garfinkle+1991} into the Kerr-Sen metric \cite{Sen1992}, both legitimate solutions of an Einstein-Maxwell-dilaton-axion theory, obtained in the low-energy effective limit of the heterotic string. Spurred by these successes, significant effort has been invested in developing and understanding such general relativistic solution-generating techniques \cite{Ernst1967, Talbot1969, Schiffer+1973, Gurses+1975, Flaherty1976, Bonnor1980, Islam1985, Giampieri1990, Quevedo1992, Drake+1997, Clement1997, Drake+1998, Stephani+2003, Glass+2004, Gibbons+2004, Ferraro2013, Keane2014, Rajan2015, Erbin2017, Contreras+2021, Casadio+2023}. 

Azreg-A{\"i}nou has recently \cite{Azreg-Ainou2014a, Azreg-Ainou2014b, Azreg-Ainou2014c} fixed one crucial source of ambiguity in the Newman-Janis algorithm%
\footnote{ In the Newman-Janis algorithm, a complex coordinate transformation involving the spin parameter, $a$, and the angular coordinate, $\vartheta$, is employed, after which a complex leg of the Newman-Penrose (NP; \cite{Newman+1965}) null tetrad ``needs fixing'' (cf. Sec. 5.1 of Ref. \cite{Rajan2015}). The inverse metric is then obtained using the fixed NP tetrad. Azreg-A{\"i}nou achieves the decomplexification of the stationary metric functions by demanding that a Boyer-Lindquist \cite{Boyer+1967} form of the metric should be attainable (cf. Sec. 7.1 of Ref. \cite{Wald1984}). This removes an ambiguity in the Newman-Janis algorithm.} %
and has obtained a stationary and axisymmetric metric that ostensibly represents a legitimate stationary generalization of an arbitrary static ``seed'' metric. Since the class of spinning Azreg-A{\"i}nou (AA) metrics generically admits a Carter constant \cite{Carter1968} for null geodesics \cite{Azreg-Ainou2014c, Shaikh2019, Kocherlakota+2023}, however, there is no guarantee that it will describe all stationary spacetimes. Furthermore, there is no concrete proof so far that this metric is the unique stationary generalization of an arbitrary static metric. Finally, the spinning AA metric is, in general, only conformally-related to its nonspinning seed metric \cite{Azreg-Ainou2014a, Azreg-Ainou2014b, Azreg-Ainou2014c}. Nevertheless, while several significant conceptual issues in this solution-generating procedure remain to be clarified, the AA algorithm has been extremely valuable in producing reasonable spinning generalizations of static regular BHs \cite{Azreg-Ainou2014c}. Here we adopt a variant of the AA metric (see Refs. \cite{Azreg-Ainou2015, Chen2022a}) which \textit{does} reduce identically (not just conformally) to the seed metric. This enables us to study the impact of spin on the properties of matter, including in the vanishing spin limit.

We begin with a general static and spherically-symmetric ``seed'' metric, given in eq. \ref{eq:Spherically_Symmetric_Metric}. Since it is always possible to describe an arbitrary static and spherically-symmetric metric using just two metric functions (see, e.g., Ch. 14 of Ref. \cite{Plebanski+2012}), it is useful to move temporarily to a different radial-coordinate $\rho$ in which the metric takes the form,
\begin{equation} \label{eq:Spherically_Symmetric_Metric_v1}
\mathrm{d}s^2 = -f\mathrm{d}t^2 + \frac{1}{f}~\mathrm{d}\rho^2 + R^2~\mathrm{d}\Omega_2^2\,,
\end{equation}
via the coordinate transformation $\mathrm{d}\rho = \sqrt{g}\mathrm{d}r$. We note that sometimes this ordinary differential equation can be solved in terms of simple analytic functions, but in other cases, it is only possible to obtain numerical solutions for $\rho(r)$ or, equivalently, for $r(\rho)$. Of the six spacetimes considered in this paper (see Table \ref{table:Known_Static_Solutions}), five can be cast analytically in the form \eqref{eq:Spherically_Symmetric_Metric_v1}. For the Frolov BH metric \cite{Frolov2016}, however, it is unlikely that an analytic form of \eqref{eq:Spherically_Symmetric_Metric_v1} exists. In equation \ref{eq:Spherically_Symmetric_Metric_v1}, the metric functions $f$ and $R$ are now functions of $\rho$ alone, i.e., $f=f(r(\rho))$ and $R=R(r(\rho))$.

We now apply the Azreg-A{\"i}nou (AA) algorithm \cite{Azreg-Ainou2014a, Azreg-Ainou2014b, Azreg-Ainou2014c} to the seed metric \eqref{eq:Spherically_Symmetric_Metric_v1} to obtain the line element, $\mathrm{d}s^2 = \mathscr{g}_{\mu^\prime\nu^\prime}\mathrm{d}x^{\mu^\prime}\mathrm{d}x^{\nu^\prime}$, of a stationary and axisymmetric metric, $\mathscr{g}_{\mu^\prime\nu^\prime}$, in Boyer-Lindquist (BL; \cite{Boyer+1967}) coordinates, $x^{\mu^\prime} = (t, \rho, \vartheta, \varphi)$, as (see also Ref. \cite{Kocherlakota+2023})
\begin{align} \label{eq:Stationary_Metric_X}
\mathrm{d}s^2
=&\ 
\frac{X}{\Sigma}\left[-\left(1-\frac{2 F}{\Sigma}\right)\mathrm{d}t^2 
-2\frac{2 F}{\Sigma}a\sin^2{\vartheta}~\mathrm{d}t\mathrm{d}\varphi \right. \\ 
&\ \quad \left.
+ \frac{\Pi}{\Sigma}\sin^2{\vartheta}~\mathrm{d}\varphi^2\nonumber + \frac{\Sigma}{\Delta}\mathrm{d}\rho^2 + \Sigma~\mathrm{d}\vartheta^2\right]\,,
\end{align}
where the spin parameter $a$ corresponds to the specific angular momentum of the spacetime. In terms of two auxiliary functions $A$ and $B$, which are related to the seed metric functions in eq. \ref{eq:Spherically_Symmetric_Metric_v1} as
\begin{equation} \label{eq:Auxiliary_Metric_Functions_v2}
A(\rho) =\ R^2(\rho)\,;\ 
B(\rho) =\ f(\rho) R^2(\rho)\,,
\end{equation}
the stationary metric functions, $F(\rho), \Delta(\rho), \Sigma(\rho, \vartheta)$, and $\Pi(\rho, \vartheta)$, are given by eq. \ref{eq:Stationary_Metric_Functions}.

The remaining metric function $X = X(\rho, \vartheta)$ can be freely prescribed. For concreteness and simplicity, we adopt the choice $X=\Sigma$. As discussed in Ref. \cite{Kocherlakota+2023}, this choice has the attractive property that if the metric \eqref{eq:Spherically_Symmetric_Metric_v1} describes an asymptotically-flat spacetime, the metric \eqref{eq:Stationary_Metric_X} is also assured to be asymptotically-flat. More importantly, this choice for the conformal factor ensures that the vanishing spin limit of the metric \eqref{eq:Stationary_Metric_X} reproduces the metric \eqref{eq:Spherically_Symmetric_Metric_v1} exactly. 

We now move to a new radial coordinate $r$ via $\mathrm{d}r = \mathrm{d}\rho/\sqrt{g}$ (this essentially inverts the previous $r\to\rho$ transformation) so that in the new set of BL coordinates, $x^\mu = (t, r, \vartheta, \varphi)$, the axisymmetric line element \eqref{eq:Stationary_Metric_X} becomes
\begin{equation} 
\begin{aligned}
\mathrm{d}s^2
=&\ 
-\left(1-\frac{2 F}{\Sigma}\right)\mathrm{d}t^2 
-2\frac{2 F}{\Sigma}a\sin^2{\vartheta}~\mathrm{d}t\mathrm{d}\varphi \\ 
&\ 
+ \frac{\Pi}{\Sigma}\sin^2{\vartheta}~\mathrm{d}\varphi^2 + \frac{\Sigma}{\Delta}g~\mathrm{d}r^2 + \Sigma~\mathrm{d}\vartheta^2\,,
\end{aligned}
\end{equation}
which matches eq. \ref{eq:Stationary_Metric}.

It is worth emphasizing that the metric \eqref{eq:Stationary_Metric} is a variant of the more popularly used AA metric (see eq. \ref{eq:AA_Metric}), and has previously also been employed in Refs. \cite{Azreg-Ainou2015, Chen2022a}. This can be seen from the zero-spin limit of the latter, which is, in general, only conformally-related to the spherically-symmetric seed metric \eqref{eq:Static_Metric}. Sec. \ref{sec:SecIII_Comoving_Frame} presents a new formal description of the background matter content associated with this geometry. A comparison to the AA algorithm framework is available in \ref{app:AppG_AA_Comparison} (to be read after Sec. \ref{sec:SecIII_Comoving_Frame}). The difference between the two frameworks becomes particularly transparent when considering the Kerr-Frolov spacetime metric introduced in Sec. \ref{sec:SecV_Spinning_BH_Properties}.


\begin{figure*}
\centering
\includegraphics[width=2\columnwidth]{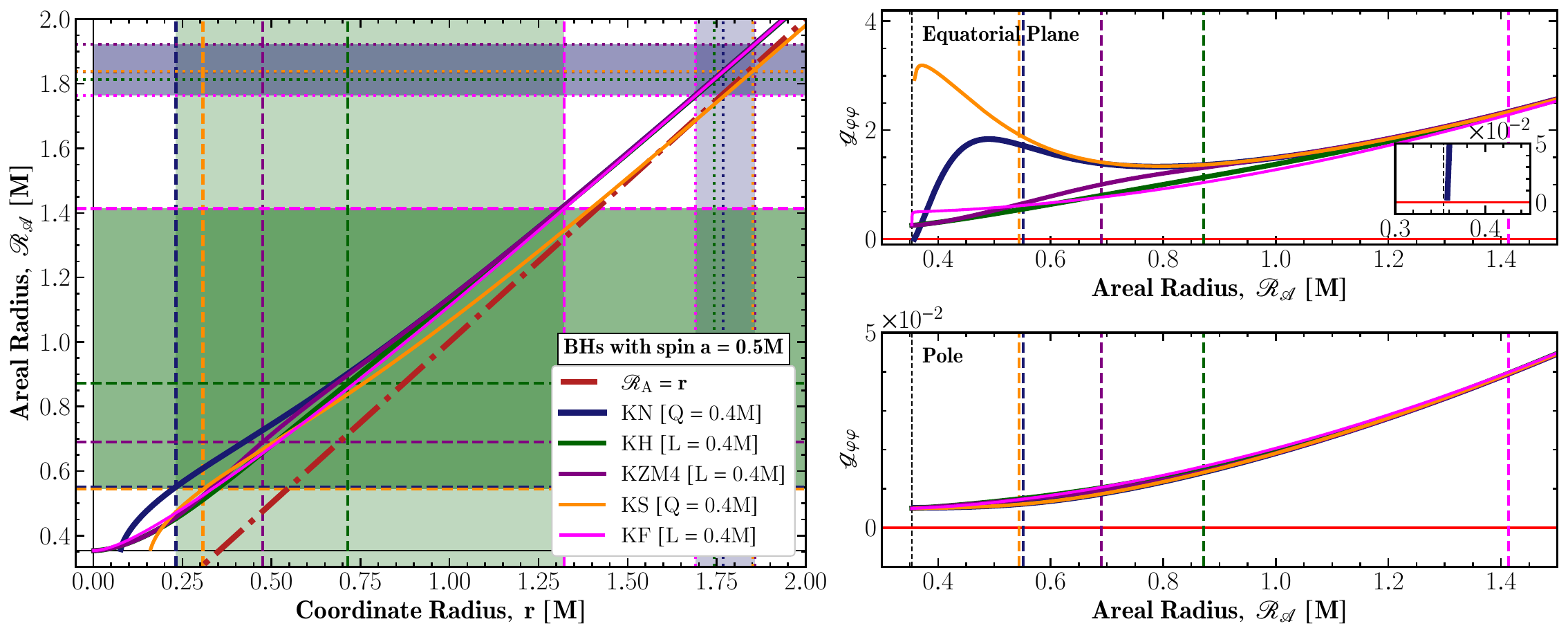}
\caption{
The left panel shows the variation of the areal radius with the coordinate radius of $2-$spheres in various spinning black hole (BH) spacetimes. The vertical and horizontal dotted lines locate the event horizons, whereas the dashed ones locate the inner horizons. The solid horizontal line locates the boundary of the region below which closed timelike curves are permitted. The panels on the right show the $\varphi\varphi-$component of the metric tensor as a function of the areal radius. For the regions considered in this work, therefore, no closed timelike curves are permitted (since $\mathscr{g}_{\varphi\varphi} > 0$).}
\label{fig:FigA1_Areal_Radius_CTCs}
\end{figure*}


\section{Areal-Radius in Axisymmetric Spacetimes and Closed Timelike Curves}
\label{app:AppB_CTCs}

The area $\mathscr{A}$ of a Boyer-Lindquist coordinate $2-$sphere of radius $r$ is given as $\mathscr{A} = \int_{\mathscr{S}}\mathrm{d}S$, where $\mathrm{d}S = \sqrt{\sigma}\mathrm{d}\vartheta\mathrm{d}\varphi$ is an area element on it, and $\sigma = \mathscr{g}_{\vartheta\vartheta}\mathscr{g}_{\varphi\varphi} = \Pi\sin^2{\vartheta}$ the determinant of its induced metric. More explicitly, we have
\begin{equation} \label{eq:Area}
\mathscr{A}(r) = 
2\pi\left[A+a^2 + \frac{\Pi(\pi/2)}{a\sqrt{\Delta}}\mathrm{arcsinh}{\left(\frac{a\sqrt{\Delta}}{\sqrt{\Pi(\pi/2)}}\right)}\right]\,.
\end{equation}
where $\Pi(\pi/2) := \Pi(r, \vartheta = \pi/2) = (A+a^2)^2 - \Delta a^2$. Equation \ref{eq:Area} is written above for both the stationary regions of spacetime ($\Delta > 0$, e.g., region I) as well as the nonstationary regions of spacetime (e.g., region II) where $\Pi > 0$. In the latter case, $r$ is a time coordinate. However, it is more intuitive to think of $r$ uniformly as the affine parameter for ingoing or outgoing null geodesics that are members of the principal null congruences (see discussion below eq. \ref{eq:PNC_inKS}) everywhere in spacetime.

The area of the event horizon, in particular, is given as,
\begin{equation}
\mathscr{A}_{\mathrm{H}} = 4\pi(A_{\mathrm{H}}+a^2)\,,
\end{equation}
where the subscript $``\mathrm{H}"$ denotes that the function is evaluated at $r=r_{\mathrm{H}}$. 

Furthermore the area of the surface $\Pi=0$ can be found to be
\begin{equation} \label{eq:CTC_Area}
\mathscr{A}_{\mathrm{CTC}} = 2\pi(A(r_{\mathrm{CTC}}) + a^2)\,.
\end{equation}
This corresponds to the area of the coordinate sphere $r=r_{\mathrm{CTC}}$ inside which closed timelike curves are permitted (since $\Pi = 0 \Rightarrow \mathscr{g}_{\varphi\varphi} = 0$). For all of the specific spacetimes considered here, barring the Kerr-Newman BH spacetime, we can set $A(r_{\mathrm{CTC}}) = 0$ (see the discussion in Sec. \ref{sec:SecVA_CTCs}). Thus, the area of the $\Pi=0$ surface is given as $\mathscr{A}_{\mathrm{CTC}} = 2\pi a^2$. 

The left panel of Fig. \ref{fig:FigA1_Areal_Radius_CTCs} shows the areal radius of coordinate $2-$spheres \eqref{eq:Areal_Radius} in all of the spinning black hole (BH) spacetimes considered here. The blue-shaded regions show the overall variation in the sizes of their event horizons and the green-shaded regions show that for their inner horizons. 

The horizontal black line on which these curves terminate bounds the region containing closed timelike curves (CTCs), which we do not consider here. 

These pathological regions of spacetime (see, e.g., \cite{Thorne1992}) occur when $\varphi\varphi$-component of the metric tensor vanishes, i.e., when $\Pi = 0$. We see from $\Pi = (A+a^2)^2 - \Delta a^2\sin^2{\vartheta}$ immediately that in regions where $\Delta < 0$, i.e., in the nonstationary cosmological regions of spacetime, $\Pi > 0$. Such regions are generically devoid of CTCs. Furthermore, it is also clear that $\Pi = 0$ must first occur in the equatorial plane. 

The right panel of Fig. \ref{fig:FigA1_Areal_Radius_CTCs} shows the variation of this metric component in the equatorial plane and on the pole, for the regions we consider here. It is verified then that no CTCs exist in these regions.


\section{Ingoing Null Coordinates}
\label{app:AppC_Null_Kerr_Schild}

In this section we introduce coordinates, $x^{\bar{\mu}} = (v, r, \vartheta, \bar{\phi})$, adapted to the ingoing PNC as (to compare with the Kerr metric, see, e.g., Sec. 5.3.6 of Ref. \cite{Poisson2004})
\begin{equation}
v = t + r_\star\,;\quad \bar{\phi} = \varphi + r_\blacklozenge\,,
\label{ingoingcoords}
\end{equation}
where%
\footnote{The new basis-vectors, $\{\partial_v, \partial_r, \partial_\vartheta, \partial_{\bar{\phi}}\}$ are given in terms of the old ones as $\{\partial_v, \partial_r, \partial_\vartheta, \partial_{\bar{\phi}}\} = \{\partial_t, \partial_r - \ell_-^t\partial_t - \ell_-^\varphi\partial_\varphi, \partial_\vartheta, \partial_\varphi\}$.}
\begin{align}
\mathrm{d}r_\star = \ell_-^t \mathrm{d}r = (A + a^2)\frac{\sqrt{g}}{\Delta}\mathrm{d}r\,; \quad
\mathrm{d}r_\blacklozenge = \ell_-^\varphi \mathrm{d}r = a\frac{\sqrt{g}}{\Delta}\mathrm{d}r\,. \nonumber
\end{align}
From the PNC generators written in these coordinates,
\begin{equation} \label{eq:PNC_inKS}
\ell_-^{\bar{\mu}} = \left[0, -1, 0, 0\right]\,,\quad \ell_+^{\bar{\mu}} = \left[(A + a^2)\frac{\sqrt{g}}{\Delta}, \frac{1}{2}, 0, a\frac{\sqrt{g}}{\Delta}\right]\,,
\end{equation}
we see that they are truly adapted to the ingoing PNC. For a closely related coordinate system adapted to a class of hypersurface-forming timelike observers, see Ref. \cite{Kocherlakota+2023}.

The above also reveals the geometric meaning of the radial coordinate $r$ in both regions of the extended spacetime (in fact, throughout the maximally-extended spacetime): Since $\mathbf{\ell_-} = -\partial_r$, it is clear that $v, \vartheta,$ and $\bar{\phi}$ are all constant for such ingoing null geodesics, and that $-r$ is the affine parameter along them (see also Sec. 5.3.6 of Ref. \cite{Poisson2004}).

In these ingoing null coordinates,%
\footnote{Outgoing null coordinates, $x^{\tilde{\mu}} = (u, r, \vartheta, \tilde{\phi})$, in which $\ell_+^{\tilde{\mu}} = +\mathbf{\delta^{\tilde{\mu}}_r}$, are arrived at by using $u=t-r_\star$ and $\tilde{\phi} = \varphi - r_\blacklozenge$ (cf. Ref. \cite{Azreg-Ainou2014c}; To compare with the Kerr metric, see, e.g., Sec. 5.6 of Ref. \cite{Hawking+1973}).
\label{fn:Outgoing_Null_Coords}} %
the metric \eqref{eq:Stationary_Metric} takes the form,
\begin{align} \label{eq:Stationary_Metric_inKS}
\mathrm{d}s^2
=&\ 
-\left(1-\frac{2 F}{\Sigma}\right)\mathrm{d}v^2 -2\frac{2 F}{\Sigma}a\sin^2{\vartheta}~\mathrm{d}v\mathrm{d}\bar{\phi} \\ 
&\ +\frac{\Pi}{\Sigma}\sin^2{\vartheta}~\mathrm{d}\bar{\phi}^2 + 
\Sigma~\mathrm{d}\vartheta^2 \nonumber \\ 
&\ + 2\sqrt{g}~\mathrm{d}v\mathrm{d}r - 2\sqrt{g}a\sin^2{\vartheta}~\mathrm{d}r\mathrm{d}\bar{\phi}\,, \nonumber
\end{align}
and has no coordinate singularity at the future event horizon, and is therefore (future) horizon-penetrating. We can use this coordinate system to describe region I, the future event horizon $\mathscr{H}^+$, as well as all of region II.

The general Killing vector, $\mathbf{K} = \mathbf{T} + \Omega \mathbf{\Phi}$, defined earliertransforms to $\mathbf{K} = \partial_v + \Omega\partial_{\bar{\phi}}$, whose norm in these coordinates, $\mathscr{g}_{\bar{\mu}\bar{\nu}}K^{\bar{\mu}}K^{\bar{\nu}}$, is precisely given by $N(\Omega)$ \eqref{eq:Killing_Norm}. Thus, there exist Killing vectors that are timelike outside the horizon, one ($\Omega = \Omega_{\mathrm{H}}$) that is null on the future horizon, but none that remain timelike in region II. 


\section{Classical Energy Conditions}
\label{app:AppD_Energy_Conditions}

The null energy condition (NEC) is the requirement that the energy density measured by an arbitrary null observer is always non-negative (see Sec. 2.1 of Ref. \cite{Poisson2004}), and is satisfied when $\epsilon + p_i \geq 0$ for each principal direction ($i$). 

The weak energy condition (WEC) requires the energy density measured by an arbitrary timelike observer to be non-negative, and is met when $\epsilon \geq0$ and $\epsilon + p_i \geq 0$. When the WEC is satisfied, the NEC is also automatically satisfied. As noted in Sec. 4.3 of Ref. \cite{Hawking+1973}, the WEC is satisfied by all experimentally detected fields. Discussed there are also the adverse consequences of WEC-violating matter.

The dominant energy condition (DEC) demands that, as measured by an arbitrary timelike observer, the energy density be non-negative, and that the flow of energy be in non-spacelike directions  (Sec. 4.3 of Ref. \cite{Hawking+1973}). This is met when $\epsilon \geq 0$ and $\epsilon \geq |p_i|$. Note that the DEC implies the WEC (and therefore the NEC). It is clear then that if the DEC is violated but the WEC is not, there exist timelike observers that view the flow of energy to be in spacelike directions. 

Finally, for completeness, we note that the strong energy condition (SEC) is satisfied if $\epsilon + p_i \geq 0$ and $\epsilon+\Sigma_ip_i \geq 0$. This last energy condition is really a statement regarding the Ricci tensor (see Sec. 2.1 of Ref. \cite{Poisson2004}), i.e., that it satisfy $R_{\mu\nu}v^\mu v^\nu \geq 0$ for an arbitrary timelike vector $v^\mu$. As discussed in Sec. 4.3 of Ref. \cite{Hawking+1973}, this ensures the ``timelike convergence condition,'' i.e., the expansion of a timelike geodesic congruence with zero vorticity will monotonically decrease. Violation of the NEC automatically implies that of the SEC. An extensive survey of various energy conditions can be found in Ref. \cite{Curiel2014}.


\section{Spherically-Symmetric Spacetimes: The Einstein Tensor and the Bianchi Identities}
\label{app:AppE_Sph_Symm_EEs}

The Einstein tensor $\hat{\mathscr{G}}$ corresponding to the static and spherically-symmetric metric \eqref{eq:Static_Metric} is diagonal and its (mixed) components are given as,
\begin{align} \label{eq:Einstein_Tensor}
\hat{\mathscr{G}}^t_{\ t} =&\ -\frac{2\partial_r m}{R^2\partial_r R}\,, \\
\hat{\mathscr{G}}^r_{\ r}-\hat{\mathscr{G}}^t_{\ t} =&\ \left(1-\frac{2m}{R}\right)\frac{2\partial_r \psi}{R\partial_r R}\,.
\end{align}
The only other nontrivial component of the Einstein tensor $\hat{\mathscr{G}}^\vartheta_{\ \vartheta} = \hat{\mathscr{G}}^\varphi_{\ \varphi}$ can be read off more conveniently from the ($r-$)Bianchi identity.
Of the Bianchi identities, $\nabla_\mu \hat{\mathscr{G}}^\mu_{\ \nu} = 0$, only the $\nu=r$ one, $0=\nabla_\mu \hat{\mathscr{G}}^\mu_{\ r} = \partial_r\hat{\mathscr{G}}^r_{\ r} + \Gamma^\mu_{\mu r}\left(\hat{\mathscr{G}}^r_{\ r} - \hat{\mathscr{G}}^\mu_{\ \mu}\right)$, is nontrivial. Here $\Gamma^\mu_{\alpha\beta}$ are the Christoffel symbols and are given as $2\Gamma^\mu_{\alpha\beta}:=\hat{\mathscr{g}}^{\mu\gamma}[-\partial_\gamma \hat{\mathscr{g}}_{\alpha\beta} + \partial_{\alpha}\hat{\mathscr{g}}_{\gamma\beta} + \partial_{\beta}\hat{\mathscr{g}}_{\alpha\gamma}]$. The $r-$Bianchi identity takes the form,
\begin{align}
0 =&\ \partial_r\hat{\mathscr{G}}^r_{\ r} + \partial_r\left[\ln{\left(R^2\sqrt{-\hat{\mathscr{g}}_{tt}}\right)}\right]\left(\hat{\mathscr{G}}^r_{\ r}-\hat{\mathscr{G}}^t_{\ t}\right) \nonumber \\ 
&\ - \partial_r\left[\ln{(R^2)}\right]\left(\hat{\mathscr{G}}^\vartheta_{\ \vartheta} - \hat{\mathscr{G}}^t_{\ t}\right)\,.
\end{align}


\section{Comoving Frame Stress-Energy Tensor for Scalar and Electromagnetic Fields}
\label{app:AppF_Matter_Models}

We turn now to the specific form of the matter stress-energy tensor for various types of matter fields, that are minimally-coupled to gravity, in their comoving frame. 


\subsection{Scalar Field}
\label{app:AppF1_Scalar_Fields}

The stress-energy tensor for a scalar field $\Phi$ with mass $m_\Phi$ that is minimally coupled to gravity is given as (see Ch. 4 of Ref. \cite{Wald1984}),
\begin{equation}
\mathscr{T}_{\mu\nu} = \nabla_\mu\Phi~\nabla_\nu\Phi - \frac{\mathscr{g}_{\mu\nu}}{2}\left[\nabla^\mu\Phi\nabla_\mu\Phi + m_\Phi^2\Phi^2\right]\,.
\end{equation}
In a static and spherically-symmetric spacetime, in particular, it is reasonable to expect that the scalar field is purely a function of $r$, i.e., $\Phi = \Phi(r)$. Thus,
\begin{align}
\hat{\mathscr{T}}_{tt} =&\ -\frac{\hat{\mathscr{g}}_{tt}}{2}\left[\hat{\mathscr{g}}^{rr}(\partial_r\Phi)^2 + m_\Phi^2\Phi^2\right]\,,\\
\hat{\mathscr{T}}_{rr} =&\ (\partial_r\Phi)^2 - \frac{\hat{\mathscr{g}}_{rr}}{2}\left[\hat{\mathscr{g}}^{rr}(\partial_r\Phi)^2 + m_\Phi^2\Phi^2\right]\,,\\
\hat{\mathscr{T}}_{\vartheta\vartheta} =&\ -\frac{\hat{\mathscr{g}}_{\vartheta\vartheta}}{2}\left[\hat{\mathscr{g}}^{rr}(\partial_r\Phi)^2 + m_\Phi^2\Phi^2\right]\,,\\
\hat{\mathscr{T}}_{\varphi\varphi} =&\ -\frac{\hat{\mathscr{g}}_{\varphi\varphi}}{2}\left[\hat{\mathscr{g}}^{rr}(\partial_r\Phi)^2 + m_\Phi^2\Phi^2\right]\,.
\end{align}
In its comoving frame \eqref{eq:Comoving_Frame_Stationary_regionI}, we can write,
\begin{align}
\hat{\mathscr{T}}_{(t)(t)} =&\ 
-\hat{\mathscr{T}}_{(\vartheta)(\vartheta)} = 
-\hat{\mathscr{T}}_{(\varphi)(\varphi)} = 
\frac{1}{2}\left[\hat{\mathscr{g}}^{rr}(\partial_r\Phi)^2 + m_\Phi^2\Phi^2\right]\,,
\nonumber \\
\hat{\mathscr{T}}_{(r)(r)} =&\ \frac{1}{2}\left[\hat{\mathscr{g}}^{rr}(\partial_r\Phi)^2 - m_\Phi^2\Phi^2\right]\,. \nonumber 
\end{align}
Finally, for a massless scalar $m_\Phi = 0$, we find that the stress-energy tensor in its comoving frame is given as, $\hat{\mathscr{T}}_{(a)(b)} = \mathrm{diag.}[\epsilon, \epsilon, -\epsilon, -\epsilon]$. 

Indeed, it can explicitly be checked that the comoving stress-energy tensor for the Wyman-Janis-Newman-Winicour \cite{Wyman1981, Janis+1968, Virbhadra1997} solution, which describes a static and spherically-symmetric configuration of a massless scalar field, is of this form.


\subsection{Electromagnetic Field}
\label{app:AppF2_EM_Fields}

The stress-energy tensor for an electromagnetic field with vector potential $\mathscr{A}_\mu$ and field strength $\mathscr{F}_{\mu\nu} = 2\nabla_{[\mu}A_{\nu]}$ that is minimally coupled to gravity is given as (see Ch. 4 of Ref. \cite{Wald1984}),
\begin{equation}
\mathscr{T}_{\mu\nu} = \frac{1}{4\pi}\left[\mathscr{g}^{\sigma\rho}\mathscr{F}_{\mu\sigma}\mathscr{F}_{\nu\rho} - \frac{\mathscr{g}_{\mu\nu}}{4}\mathscr{F}_{\sigma\rho}\mathscr{F}^{\sigma\rho}\right]\,.
\end{equation}
For an electric field in a static and spherically-symmetric spacetime, in particular, we can write $\hat{A}_\mu = [\Phi(r), 0, 0, 0]$. Here $\Phi(r)$ is the Coulomb potential for this configuration. Thus, $\hat{\mathscr{F}}_{\mu\nu} = 2(\partial_r\Phi) \delta_{[\mu}^r\delta_{\nu]}^t$ and $\hat{\mathscr{F}}_{\mu\nu}\hat{\mathscr{F}}^{\mu\nu} = 2(\partial_r\Phi)^2\hat{\mathscr{g}}^{tt}\hat{\mathscr{g}}^{rr}$. With this, we can write the stress-energy tensor as,
\begin{equation}
\hat{\mathscr{T}}_{\mu\nu} = \frac{(\partial_r\Phi)^2}{4\pi}\left[\hat{\mathscr{g}}^{rr}\delta^t_\mu\delta^t_\nu + \hat{\mathscr{g}}^{tt}\delta^r_\mu\delta^r_\nu - \frac{\hat{\mathscr{g}}_{\mu\nu}}{2}\hat{\mathscr{g}}^{tt}\hat{\mathscr{g}}^{rr}\right]\,.
\end{equation}
In its comoving frame \eqref{eq:Comoving_Frame_Stationary_regionI}, we can write,
\begin{equation}
\hat{\mathscr{T}}_{(t)(t)} = -\hat{\mathscr{T}}_{(r)(r)} = \hat{\mathscr{T}}_{(\vartheta)(\vartheta)} = \hat{\mathscr{T}}_{(\varphi)(\varphi)} = -\frac{(\partial_r\Phi)^2}{8\pi}\hat{\mathscr{g}}^{tt}\hat{\mathscr{g}}^{rr}\,. \nonumber
\end{equation}
Thus, the comoving frame stress-energy tensor for a static and spherically-symmetric configuration of electric field is of the form $\hat{\mathscr{T}}_{(a)(b)} = \mathrm{diag.}[\epsilon, -\epsilon, \epsilon, \epsilon]$. 

Indeed, the vector potential for the electrically-charged  Reissner-Nordstr{\"o}m (RN; \cite{Wald1984}) solution is given as $\hat{A}_\mu = (-Q/r, 0, 0, 0)$ so that $\hat{\mathscr{T}}_{(a)(b)} = Q^2/(8\pi r^4)\mathrm{diag.}[1, -1, 1, 1]$

For a magnetic field in a static and spherically-symmetric spacetime, in particular, we can write $\hat{A}_\mu = [0, 0, 0, \Psi(\vartheta)]$. Here $\Psi(\vartheta)$ is related to the poloidal magnetic flux function (see, e.g., Ref. \cite{Narayan+2006}) for this configuration. Thus, $\hat{\mathscr{F}}_{\mu\nu} = 2(\partial_\vartheta\Psi) \delta_{[\mu}^\vartheta\delta_{\nu]}^\varphi$ and $\hat{\mathscr{F}}_{\mu\nu}\hat{\mathscr{F}}^{\mu\nu} = 2(\partial_\vartheta\Psi)^2\hat{\mathscr{g}}^{\vartheta\vartheta}\hat{\mathscr{g}}^{\varphi\varphi}$. With this, we can write the stress-energy tensor as,
\begin{equation}
\hat{\mathscr{T}}_{\mu\nu} = \frac{(\partial_\vartheta\Psi)^2}{4\pi}\left[\hat{\mathscr{g}}^{\varphi\varphi}\delta^\vartheta_\mu\delta^\vartheta_\nu + \hat{\mathscr{g}}^{\vartheta\vartheta}\delta^\varphi_\mu\delta^\varphi_\nu - \frac{\hat{\mathscr{g}}_{\mu\nu}}{2}\hat{\mathscr{g}}^{\vartheta\vartheta}\hat{\mathscr{g}}^{\varphi\varphi}\right]\,. \nonumber
\end{equation}
In its comoving frame \eqref{eq:Comoving_Frame_Stationary_regionI}, we can write,
\begin{align}
&\ \hat{\mathscr{T}}_{(t)(t)} = -\hat{\mathscr{T}}_{(r)(r)} = \hat{\mathscr{T}}_{(\vartheta)(\vartheta)} = \hat{\mathscr{T}}_{(\varphi)(\varphi)} \nonumber \\
&\ = +\frac{(\partial_\vartheta\Psi)^2}{8\pi}\hat{\mathscr{g}}^{\vartheta\vartheta}\hat{\mathscr{g}}^{\varphi\varphi}\,. \nonumber
\end{align}
Thus, the comoving frame stress-energy tensor for a static and spherically-symmetric configuration of a magnetic field is also of the form $\hat{\mathscr{T}}_{(a)(b)} = \mathrm{diag.}[\epsilon, -\epsilon, \epsilon, \epsilon]$. 

Indeed, the vector potential for the magnetically-charged RN solution is given as $\hat{A}_\mu = (0, 0, 0, Q\cos{\vartheta})$. 


\section{Comparison to the Azreg-A{\"i}nou Algorithm}
\label{app:AppG_AA_Comparison}

Applying the Azreg-A{\"i}nou (AA) algorithm \cite{Azreg-Ainou2014a, Azreg-Ainou2014b, Azreg-Ainou2014c} directly on the general spherically-symmetric seed metric \eqref{eq:Spherically_Symmetric_Metric} leads to an axisymmetric metric, which we refer to as the AA metric, of the form (cf. eq. 2 of Ref. \cite{Kocherlakota+2023}),
\begin{align} \label{eq:AA_Metric}
\mathrm{d}s^2
=&\ 
\frac{X}{\Sigma}\left[-\left(1-\frac{2 F}{\Sigma}\right)\mathrm{d}t^2 
-2\frac{2 F}{\Sigma}a\sin^2{\vartheta}~\mathrm{d}t\mathrm{d}\varphi\right. \\ 
&\ 
\left. +\frac{\Pi}{\Sigma}\sin^2{\vartheta}~\mathrm{d}\varphi^2 + \frac{\Sigma}{\Delta}\mathrm{d}r^2 + \Sigma~\mathrm{d}\vartheta^2\right]\,, \nonumber
\end{align}
where the metric functions, $F, \Delta, \Sigma, \Pi$, are expressed in terms of auxiliary functions $A$ and $B$, as in eq. \ref{eq:Stationary_Metric_Functions}. However, it is important to note that the auxiliary functions for the AA metric (see eq. 3 of Ref. \cite{Kocherlakota+2023}),
\begin{equation} \label{eq:Auxiliary_Metric_Functions_AA}
A(r) = R^2/\sqrt{g}\,;\ B(r) = (f/g)R^2\,,
\end{equation}
differ from the $A(r)$, $B(r)$ we define in equation \eqref{eq:Auxiliary_Metric_Functions} for the \textit{ansatz} metric employed here. The two metrics (\ref{eq:Stationary_Metric} and \ref{eq:AA_Metric}) also differ, as written, additionally in the form of their $rr-$components. Finally, the conformal factor $X$ (denoted by $\Psi$ in Refs. \cite{Azreg-Ainou2014a, Azreg-Ainou2014b, Azreg-Ainou2014c}) in the AA metric \eqref{eq:AA_Metric} is, in general, an unknown.

The AA algorithm \cite{Azreg-Ainou2014a, Azreg-Ainou2014b, Azreg-Ainou2014c} provides a prescription to fix the conformal factor $X$ by solving a nonlinear hyperbolic partial differential equation (PDE) and a linear PDE. These equations arise from demanding that the rest-frame of the matter generating the spacetime \eqref{eq:AA_Metric} be of the form \eqref{eq:Einstein_Eigenvectors}, with $\Omega = \Omega_{\mathrm{PNC}} = a/(A+a^2)$. Tetrad orthonormality (see eq. \ref{eq:Chi}) automatically fixes $\chi = 1/(a\sin^2{\vartheta})$.

The first of the two PDEs is a concomitant of demanding that the $r\vartheta-$component of the Einstein tensor of the AA metric \eqref{eq:AA_Metric} vanish, i.e., $\mathscr{G}_{r\vartheta} = 0$,%
\footnote{Note that demanding that $e_{(r)} \propto \partial_r$ and $e_{(\vartheta)} \propto \partial_\vartheta$ when $X\neq\Sigma$ is significantly stronger than requiring that $\mathscr{G}_{(r)(\vartheta)} = 0$ (cf. Footnote \ref{fn:X_neq_Sigma}).
} %
and the second is a consequence of demanding that $\mathscr{G}_{(t)(\varphi)} = 0$ for $\Omega=\Omega_{\mathrm{PNC}}$ specifically. These PDEs are given respectively as (with $y=\cos{\vartheta}$; cf. eq. 4 and 7 of Ref. \cite{Azreg-Ainou2014a}),
\begin{align} 
\label{eq:AA_PDE_1}
0 =&\ \sqrt{X}\partial_r\partial_y\left(1/\sqrt{X}\right) - \sqrt{\Sigma}\partial_r\partial_y\left(1/\sqrt{\Sigma}\right) \\
\label{eq:AA_PDE_2}
0 =&\ X\left[-\Sigma\partial_r^2A + (\partial_r A)^2 + 2(A - a^2y^2)\right] \\ 
&\ + \Sigma\left[2y\partial_yX - \partial_r A \cdot \partial_r X\right]\,. \nonumber 
\end{align}
While $X=\Sigma$ is clearly a (trivial) solution of \eqref{eq:AA_PDE_1}, there is no guarantee that it also solves \eqref{eq:AA_PDE_2}. This is of critical importance since for a solution $X \neq \Sigma$, the zero-spin limit of the AA metric \eqref{eq:AA_Metric} is, in general, only conformally-related (by a factor of $\sqrt{g(r)}$) to the spherically-symmetric seed metric \eqref{eq:Spherically_Symmetric_Metric} that was used as an input (see Sec. 4 of Ref. \cite{Azreg-Ainou2014a}). 

Remembering that $\Sigma = A+a^2y^2$, we recognize that both PDEs involve only the metric functions $X$ and $A$. Setting $X=\Sigma$ in eq. \ref{eq:AA_PDE_2} yields an ordinary differential equation for $A(r)$: $\partial_r^2 A = 2$. Thus, the choice $X=\Sigma$ yields a legitimate spinning AA counterpart for seed metrics with $A(r) = c_2 r^2 + c_1 r + c_0$, where $c_i$ are constants (cf. eq. 8 of Ref. \cite{Azreg-Ainou2014a}). Since $A=R^2/\sqrt{g}$ for the AA metric \eqref{eq:Auxiliary_Metric_Functions_AA}, the equation above does not imply specific forms for either $R$ or $g$. However, for the special case when the seed metric is of the form
\begin{equation} \label{eq:PNC_Metric_Class}
R^2(r) = c_2 r^2 + c_1 r + c_0\,;\ \  g(r) = 1\,,
\end{equation}
we find that the zero-spin limit of the AA metric identically matches the seed metric. For other choices of $R$ and $g$ (\textit{even when} $X=\Sigma$), the former is, in general, only conformally-related to the latter. The GMGHS metric \cite{Gibbons+1988, Garfinkle+1991} is an example of a seed metric that satisfies eq. \ref{eq:PNC_Metric_Class} (see Table \ref{table:Known_Static_Solutions}). Note that the above contains the important class of degenerate spacetimes (Sec. \ref{sec:SecIIIB_Degenerate_Spacetimes}), i.e., $c_2 = 1, c_1 = c_0 = 0$ (cf. Sec. 3 of Ref. \cite{Azreg-Ainou2014a}). 

We have already alluded to the differences between the \textit{ansatz} metric described here \eqref{eq:Stationary_Metric} and the AA metric \eqref{eq:AA_Metric}, as well as between the methods used to obtain the two metrics.  We now highlight two key points. First, we apply the AA algorithm to a general spherically-symmetric metric \textit{in specific coordinates} \eqref{eq:Spherically_Symmetric_Metric_v1}. Second, we \textit{impose} $X=\Sigma$, ensuring that we recover the spherically-symmetric seed metric in the zero-spin limit of its stationary generalization \eqref{eq:Stationary_Metric}, and then solve for the matter rest-frame via the procedure described in Sec. \ref{sec:SecIII_Comoving_Frame} instead. As noted above, the latter reduces to solving a single quadratic equation \eqref{eq:Omega_Quadratic} for $\Omega$, allowing us to bypass the nontrivial PDE system (\ref{eq:AA_PDE_1}, \ref{eq:AA_PDE_2}). In particular, we do not require the matter rest frame to satisfy $\Omega = \Omega_{\mathrm{PNC}}$.

In conclusion, we emphasize that both methods can be used to obtain legitimate spinning counterparts when starting from the same seed metric. For spherically-symmetric seed metrics with metric functions \eqref{eq:PNC_Metric_Class}, both methods output the same axisymmetric metric. This is because (i) when $g(r)=1$ both auxiliary metric functions, $A$ and $B$, in both methods (\ref{eq:Auxiliary_Metric_Functions_v2}, \ref{eq:Auxiliary_Metric_Functions_AA}) match identically, and (ii) we find $\Omega = \Omega_{\mathrm{PNC}}$ (which the AA method imposes on its solution) to be the unique legitimate solution \eqref{eq:Omega_PNC} to the quadratic equation for the matter rest-frame angular velocity in our framework, and $X=\Sigma$ (which we impose right from the start) is a legitimate solution to the PDE system in the AA framework. However, since we do not impose the requirement $\Omega = \Omega_{\mathrm{PNC}}$ in our framework, it is possible for matter in spacetimes described by \eqref{eq:Stationary_Metric} to be differentially-rotating on a coordinate $2-$sphere (see the Kerr-Frolov spacetime below; Sec. \ref{sec:SecVC_Non_Rigidly_Rotating}), which is not permitted in the Azreg-A{\"i}nou framework (those solutions have matter that is always rotating rigidly in spherical shells). Therefore, while our method relies crucially on the AA algorithm, it provides an alternative implementation of, and a distinct new perspective on, this thread of solution-generating techniques.


\bibliography{Refs-Matter-in-Stationary-Spacetimes.bib}


\end{document}